\documentclass[12pt,preprint]{aastex}
\usepackage{lscape}

\accepted{}
\journalid{}{}
\articleid{}{}

\shortauthors{Mohanty et al..}
\shorttitle{On The Planetary Mass Status of 2M1207B}

\received{2005 October}
\begin{document}

\def\lam{$\lambda$}
\def\tross{${\tau}_{R}$ }
\def\hal{{{\rm H}\alpha} }
\def\rc{R_C}
\def\ic{I_C}
\def\rcm{R_{Cm}}
\def\icm{I_{Cm}}
\def\jm{J_{m}}
\def\rco{R_{Co}}
\def\ico{I_{Co}}
\def\jo{J_{o}}
\def\av{A_V}
\def\ar{A_{R_C}}
\def\ai{A_{I_C}}
\def\exj{A_{J}}
\def\kr{k_{R_C}}
\def\ki{k_{I_C}}
\def\rad{{\mathcal{R}}_{\ast}}
\def\dist{\mathcal{D}}
\def\mass{{\mathcal{M}}_{\ast}}
\def\lum{{\mathcal{L}}_{bol}}
\def\logten{log_{10}}

\def\li{\ion{Li}{1}\ }
\def\na{\ion{Na}{1}\ }
\def\pot{\ion{K}{1}\ }
\def\oxy{\ion{O}{1}\ }
\def\hel{\ion{He}{1}\ }
\def\cal{\ion{Ca}{2}\ }
\def\nit{\ion{N}{2}\ }
\def\sul{\ion{S}{2}\ }

\def\ualph{${\mu}_{\alpha}$ }
\def\udelt{${\mu}_{\delta}$ }
\def\kms{km s$^{-1}$\ }
\def\kmsp{km s$^{-1}$ pix$^{-1}$\ }
\def\cms{cm s$^{-1}$\ }
\def\cmc{cm$^{-3}$\ }
\def\cmss{cm$^{2}$ s$^{-1}$\ }
\def\cmcs{cm$^{3}$ s$^{-1}$\ }

\def\mdot{$\dot{M}$ }
\def\msun{M$_\odot$ }
\def\rsun{R$_\odot$ }
\def\lsun{L$_\odot$ }
\def\mj{M$_{Jup}$ }

\def\teff{T$_{e\! f\! f}$~}
\def\tefflbol{T$_{e\! f\! f}$-L$_{\it bol}$~}
\def\gv{{\it g}~}
\def\vsini{{\it v}~sin{\it i}~}
\def\vrad{v$_{\it rad}$~}
\def\lbol{L$_{\it bol}$~}
\def\mbol{{\rm M}_{\it bol}}
\def\lhal{L_{H\alpha}}
\def\fhal{F_{H\alpha}}
\def\fcal{{\mathcal{F}}_{CaII}}
\def\fcont{{\mathcal{F}}_{cont}}
\def\fbol{F_{\it bol}}
\def\lx{L_{X} }
\def\eqwhal{EW_{H\alpha}}
\def\eqwcal{EW_{CaII}}
\def\alom{{\alpha}{\Omega} }
\def\ross{R_{0} }
\def\cots{{\tau}_{c} }
\def\fchal{{\mathcal{F}}_{c\hal} }
\def\h2o{H$_2$O}
\def\vcal{r_{CaII} }
\def\fex{{\mathcal{F}}_{excess}}

\title{The Planetary Mass Companion 2MASS1207-3932 B: \\Temperature, Mass and Evidence for an Edge-On Disk}

\author{Subhanjoy Mohanty\altaffilmark{1},  Ray Jayawardhana\altaffilmark{2}, Nuria Hu\'{e}lamo\altaffilmark{3}, Eric Mamajek\altaffilmark{4}}

\altaffiltext{1}{Harvard-Smithsonian Center for Astrophysics, Cambridge, MA 02138, USA.  smohanty@cfa.harvard.edu}
\altaffiltext{2}{Department of Astronomy \& Astrophysics, University of Toronto, Toronto, ON M5S 3H8, CANADA.  rayjay@astro.utoronto.ca}
\altaffiltext{3}{VLT, European Southern Observatory, CHILE.  nhuelamo@eso.org}
\altaffiltext{4}{Harvard-Smithsonian Center for Astrophysics, Cambridge, MA 02138, USA.  emamajek@cfa.harvard.edu}

\begin{abstract} 
We present $J$--band imaging and $H$+$K$--band low-resolution spectroscopy of the 2MASS1207-3932AB system, obtained with the VLT NIR-AO instrument (NACO).  Our $J$-band astrometry is consistent with AB being a co-moving system, in agreement with the recent results of Chauvin et al. (2005a).  For the putative planetary mass secondary, we find $J$ = 20.0$\pm$0.2 mag.  The shapes of the $HK$ spectra for both components imply low gravity, as expected for this young (5--10 Myr-old) system, as well as a dusty atmosphere for the secondary.  Comparisons to the latest synthetic spectra yield {T$_{eff-A}$} $\approx$ 2550$\pm$150K, and {T$_{eff-B}$} $\approx$ 1600$\pm$100K.  These temperatures are consistent with the late-M and mid-to-late L spectral types derived earlier for 2M1207A and B respectively.  For these \teff, and an age of 5--10 Myrs, the latest theoretical evolutionary tracks imply $M_A$ $\approx$ 24$\pm$6 M$_{Jup}$ and $M_B$ $\approx$ 8$\pm$2 M$_{Jup}$. Independent comparisons of the theoretical tracks to the observed colors, spanning $\sim$$I$ to $L'$ (including recent HST photometry), also yield the same mass and temperature estimates.  Our mass for the primary agrees with other recent analyses; however, our secondary mass, while still in the planetary regime, is 2--3 times larger than claimed previously.  The roots of this discrepancy can be traced directly to the luminosities: while the absolute photometry and $\mbol$ of the primary are in excellent agreement with theoretical predictions (especially with the recently derived $d$ = 53$\pm$6 pc for the system), the secondary appears $\sim$ 2.5$\pm$0.5 mag {\it fainter} than expected in all photometric bands from $I$ to $L'$ as well as in $\mbol$. This anomalous under-luminosity accounts for the much lower secondary mass (and temperature) derived in earlier studies.  We argue that this effect is highly unlikely to result from: {\it (i)} a large overestimation of our secondary \teff;  {\it (ii)} serious overestimation of luminosities by the theoretical evolutionary models; {\it (iii)} very large distance/age variations between the two components; or {\it (iv)} faintness in the secondary due to formation via core-accretion. These conclusions are bolstered by the absence of any luminosity problems with the primary in our analysis.  Similarly, we find no luminosity discrepancies in the recently discovered sub-stellar companion AB Pic B, which is also young (age $\sim$30 Myr) and comparable in spectral classification ($\sim$L-type) and temperature ($\sim$1700K) to 2M1207B.  We therefore suggest grey extinction in 2M1207B, due to occlusion by an edge-on circum-secondary disk.  This scenario is consistent with the observed properties of edge-on disks around T Tauri stars, and with the known presence of a high-inclination evolved disk around the primary.  Finally, the system's implied mass ratio of $\sim$0.3 suggests a binary-like formation scenario.
  
\end{abstract}

\keywords{stars: low-mass, brown dwarfs -- stars: pre-main sequence -- stars: formation -- circumstellar matter -- planetary systems -- techniques: spectroscopic} 

\section{Introduction}
The nearby young brown dwarf 2MASS J1207334-393254 (hereafter 2M1207A) has been the subject of widespread attention in recent years. Based on its location, proper motion and spectral signatures of youth, Gizis (2002) proposed that it is a likely member of the $\sim$8-Myr-old TW Hydrae Association (henceforth TWA), one of the closest known young clusters ($\sim$50 pc). His low-resolution optical spectrum also evinced strong $\hal$ emission, possibly indicative of disk accretion. High-resolution optical spectra subsequently obtained by Mohanty et al. (2003) showed the $\hal$ emission profile to be broad, asymmetric and variable over timescales of hours, and also revealed strong emission in many other Balmer lines and in HeI.  These characteristics greatly bolstered the case for ongoing accretion.  Mohanty et al. further derived radial and rotational velocities for 2M1207A, and showed the radial velocity to be congruent with that expected of a TWA member.  Finally, the X-ray versus $\hal$ analysis by Gizis \& Bharat (2004) provided additional support for ongoing disk accretion in 2M1207A.  

Mohanty et al. (2005) subsequently calculated an accretion rate of approximately $<$10$^{-11}$ \msun{yr$^{-1}$}.  More recently, Scholz et al. (2005) and Scholz \& Jayawardhana (2006) have found dramatic changes in the shape and intensity of 2M1207A's H$\alpha$ profile over timescales ranging from hours to weeks, indicating variations in the accretion rate by a factor of 5-10 over $\sim$six weeks (with the lower rates comparable to Mohanty et al.'s (2005) value).  These authors have also identified a red-shifted absorption component in $\hal$, quasi-periodic over day-long timescales.  The presence of this component indicates asymmetric, magnetically channeled accretion columns rising from the surrounding disk; furthermore, its $\sim$1 day periodicity, compared to the \vsini found by Mohanty et al. (2003) and the expected radii of young brown dwarfs, suggests that the disk is viewed at a high inclination angle (i.e., relatively close to edge-on; $i$ $\gtrsim$ 60$^{\circ}$).  

Jayawardhana et al. (2003) did not find any $L'$-band (3.8$\mu$m) excess disk emission from 2M1207A.  However, Sterzik et al. (2004) detected $\sim$8--11$\mu$m mid-infrared excess from the disk, a situation similar to two other stellar members of TWA (Jayawardhana et al. 1999).  Gizis et al. (2005b) have recently detected the disk out to 24$\mu$m with $Spitzer$.  The shape of the mid-IR spectral energy distribution (SED), combined with the lack of 10$\mu$m silicate emission (Sterzik et al. 2004), suggests that the dust disk is flat, with significant grain growth and/or settling.  Warm disk gas has now also been seen around 2M1207A, via the detection of fluorescent H$_2$ emission (Gizis et al. 2005a).  Together, all these data establish 2M1207A as the {\it closest known young brown dwarf with a surrounding accretion disk}.  

Meanwhile, Chauvin et al. (2004) announced the direct detection of a possibly planetary mass companion to 2M1207A, at a projected separation of $\sim$0.78\arcsec.  Their $H$$K_s$$L'$ photometry suggested a spectral type of $\sim$L5--L9.5 for the secondary, and their $H$-band low-resolution spectrum, while quite poor in S/N, appeared consistent with this estimate.  HST NICMOS photometry by Song et al. (2006), covering roughly $I$ to $H$ bands, also indicates a similar spectral type (mid-to-late L).  Assuming a distance of $\sim$70 pc, Chauvin et al. (2004) derived a luminosity for the secondary; adopting an age of $\sim$8 Myr, and comparing to age-luminosity evolutionary models, they then arrived at a mass of $\sim$5 \mj and effective temperature (T$_{eff}$) of $\sim$1250K.  The following year, Chauvin et al. (2005a) presented astrometric observations that strongly supported the physical association of the primary (2M1207A) and secondary (2M1207B): the two almost certainly form a single, bound, co-moving system.

Subsequently, Mamajek (2005), with refined space motions for 2M1207A and TWA combined with a detailed moving-cluster analysis, showed that: {\it (1)} 2M1207A is indeed a bona-fide member of TWA, and {\it (2)} its distance is 53$\pm$6 pc, in agreement with the known distances to other TWA members and significantly less than the 70 pc assumed by Chauvin et al. (2004, 2005a).  Song et al. (2006) have also arrived at a very similar distance, using a scaled proper motion technique: 59$\pm$7 pc, within 1$\sigma$ of Mamajek's value and again lower than Chauvin et al.'s estimate.  The reduction in distance decreases the estimated luminosity of 2M1207B, and thus lowers its inferred mass: employing his new distance, but otherwise carrying out exactly the same age-luminosity analysis as Chauvin et al. (2004), Mamajek (2005) derived a mass of just 3--4 \mj for the secondary.  While Mamajek did not state what \teff this mass corresponds to, the evolutionary models he uses (Chabrier et al. 2000; Baraffe et al. 2003) indicate \teff $\sim$900--1050K, at the adopted age of $\sim$8 Myr.  

2M1207B is thus unique in several respects: it is the coolest known confirmed substellar object at an age of a few Myrs; if the above mass estimates are valid, it is also arguably the first directly imaged planetary mass companion outside the Solar System (and certainly the first imaged one with confirmation of common proper motion), and the first such body discovered around a brown dwarf.  From the point of view of substellar binary systems too, 2M1207AB is quite remarkable.  On the one hand, given the high multiplicity frequency among low mass stars in TWA (Brandeker et al. 2003), it is perhaps not surprising to find a substellar binary like 2M1207AB. However, brown dwarf binaries wider than 20 AU appear to be absent in older ($\sim$100 Myr-old) open clusters (e.g., Mart\'{i}n et al. 2003; Bouy et al. 2006a), as well as in the nearby field (Bouy et al. 2003; Burgasser et al. 2003a).  In very young ($\sim$1--5 Myr-old) star-forming regions too, the substellar binary separation seems comparable to that in the field (Kraus et al. 2005, 2006; although there is a hint that some wider brown dwarf binaries may exist at such ages; Bouy et al. 2006b; Luhman 2005).  With a separation of at least $\sim$40 AU (using the projected angular separation and the Mamajek 2005 distance), the $\sim$8 Myr-old 2M1207AB system is definitely an outlier compared to these results, and can potentially illuminate evolutionary and/or environmental effects on substellar binarity.   In short, 2M1207B offers critical insights into the properties of ultra-cool objects at young ages, the origin of planetary mass companions, the formation of substellar bodies in general, and binarity in the brown dwarf domain.  

\section{Spectral Type versus Temperature Discrepancy in Earlier Analyses}

However, one fundamental aspect of 2M1207B's inferred properties is rather troubling.  The \teff $\sim$ 1250K derived by Chauvin et al. (2004), based ultimately on an over-large distance, corresponds in the field to early T dwarfs (Golimowski et al. 2004).  The $\sim$900--1050K implied by Mamajek's (2005) analysis, based on a more rigorous distance estimate, corresponds to mid-to-late T dwarfs.  Finally, the \teff $\sim$ 1250K derived by Song et al. (2006) from HST absolute photometry down to $\sim$$I$ again corresponds, like Chauvin et al.'s estimate, to early T dwarfs.   Thus, at least compared to field dwarfs, even Chauvin et al.'s (2004) and Song et al.'s (2006) \teff is at odds with the mid-to-late L type they themselves observe for 2M1207B, while the \teff suggested by Mamajek's (2005) analysis is entirely inconsistent with this spectral type.  Specifically, the settling of photospheric dust, and the appearance of CH$_4$ absorption in the $K$-band, already make early T dwarfs {\it bluer} in $H$-$K$ than hotter, earlier spectral types, while mid-to-late T dwarfs are severely bluer (e.g., Knapp et al. 2004).  In contrast, Chauvin et al. (2004) show that 2M1207B is extremely {\it red} in $H$-$K$ (which is why they derive a mid-to-late L type in the first place), suggesting a significantly higher \teff than inferred from their, Song et al.'s or Mamajek's luminosity analyses.  In short, there seems to be a serious discrepancy between the \teff implied for 2M1207B by its estimated luminosity, and the \teff suggested by its observed spectral type.  

Of course, 2M1207B is much younger than field dwarfs, so this apparent discrepancy may be due to its much lower gravity.  Either way, given the importance of this object, the mismatch is worth investigating in detail.  To this end, we have obtained $J$-band photometry and $H$$K$-band low-resolution spectroscopy of both 2M1207A and B on the European Southern Observatory's Very Large Telescope (VLT), and compared our data (along with previous photometry) to the latest synthetic spectra and evolutionary models. Our spectral and color analyses directly yield temperature and mass estimates, independent of luminosity.  The \teff we derive for both bodies are consistent with their previously cited spectral types.  In particular, the \teff implied by 2M1207B's spectral energy distribution is substantially higher than suggested by Chauvin et al.'s (2004), Song et al.'s (2006) and Mamajek's (2005) luminosity considerations, as suspected a priori from the spectral type deliberations above.  Low gravity alone cannot resolve this discrepancy.  Instead, it appears that 2M1207B is anomalously underluminous, due to grey extinction in the infrared.  We support this conclusion by showing that the recently discovered young brown dwarf AB Pic B (Chauvin et al. 2005b), which is roughly similar to 2M1207B in temperature, spectral type and age, does {\it not} exhibit any temperature-luminosity discrepancies: 2M1207B is indeed peculiarly faint. We ascribe this to the presence of a nearly edge-on disk around 2M1207B.  Our mass for 2M1207B is also correspondingly higher (though still in the planetary regime), supporting the idea that the 2M1207AB system formed in a manner analogous to stellar binaries.  

\section{Observations}
$J$-band imaging and $H$+$K$-band low-resolution spectroscopy of 2M1207A and B were performed on ESO's VLT in Chile, using the Near-infrared Adaptive Optics System (NAOS) with the infrared camera and spectrometer CONICA (NACO).  Wavefront sensing was in the near-infrared, guiding on the primary (2M1207A).  The N90C10 dichroic (which sends 90\% of the light to the wavefront-sensor and 10\% to the science camera), High Sensitivity detector mode, and Fowler Nsample readout mode were employed throughout.  Observations were in Service Mode, and constrained to target airmass $<$1.2 and visual seeing (before AO-correction) $\lesssim$ 0.6\arcsec~.       

The $J$-band images were acquired on March 26 2005, with the CONICA S27 camera (pixel-scale $\approx$ 27 mas/pix, FOV = 28\arcsec$\times$28\arcsec).  80 images were taken, each of duration 30s, yielding a total integration time of 40 min.
   
The $H$+$K$-band low-resolution spectroscopy was carried out over the period April to June 2005, with the $HK$-filter (simultaneous coverage over $H$ and $K$ bands, 1.3--2.6$\mu$m), a slit-width of 86 mas, and the S54 camera (pixel-scale $\approx$ 54 mas/pix, FOV = 56\arcsec$\times$56\arcsec).  The spectral resolution was $\sim$550. 31 spectra were obtained, in 4 observing blocks (OBs) of 8 spectra each (except for one June 2005 OB, in which 7 spectra were taken instead of 8; see below); exposure time for each frame was 300s, giving a total integration time of 155 min. 
 
The 8 spectra in each OB were taken in an `abbaabba' nod-cycle, i.e., as 4 pairs of spectra at 2 nod positions each.  The slit was oriented along the 2M1207A--B axis, with nodding along the slit, yielding simultaneous spectra of both components at each nod.  A glitch terminated the first OB of June 3 2005 after only 7 nods, before the last spectrum in the cycle was taken; a second complete block of 8 spectra was therefore obtained on the same night.  Subsequent examination showed the 7 spectra from the first OB to be perfectly viable as well, and they are included in our analysis along with all the other spectra.

Finally, we also use previously published photometry of 2M1207 in our analysis: ground-based $J$, $H$, $K_s$, $L'$ for the primary (2MASS All Sky Catalog of Point Sources, Cutri et al. 2003; Jayawardhana et al. 2003) and $H$, $K_s$, $L'$ for the secondary (Chauvin et al. 2004).  

\section{Data Reduction}  

\subsection{Photometry}
The $J$-band images were flat-fielded, registered and co-added using the NACO-pipeline reduction software ECLIPSE (Devillard 1997).  2M1207A and B are clearly visible as separate sources; a background K-type star is also visible south-east of the 2M1207 system (Figs. 1 and 2).  Apparent $J$ magnitudes for both 2M1207A and the K-type star are available in the 2MASS database\footnote{While the 2MASS spatial resolution is too low to distinguish between 2M1207A and B, B alone is well below the survey's detection limit, so the 2MASS fluxes for 2M1207 correspond to that of A alone.}.  The $J$ magnitude of 2M1207B was determined through relative photometry with respect to A; we checked the validity of our flux calculations by similarly deriving the $J$ magnitude of the K-type star as well, relative to 2M1207A, and comparing to its 2MASS value.

Background-subtracted counts (in arbitrary units) for 2M1207A and the K-type star were determined first, with the $IRAF$ $apphot$ package.  Our errors in photometry are dominated by photon (shot) noise.  While 2M1207B sits on the wings of the primary, its flux contribution relative to the primary is completely negligible ($\ll$1\%, see below), and no attempt was made to exclude it while extracting the counts for 2M1207A.  Adopting the 2MASS magnitude for 2M1207A ($J$ = 13.00$\pm$0.03), and the magnitude difference between it and the (fainter) K-type star implied by our aperture photometry ($\Delta$$J$ = 2.32$\pm$0.01), yields $J$ = 15.32$\pm$0.03 for the latter.  This agrees very well with the 2MASS magnitude for this source, $J$ = 15.39$\pm$0.07, supporting the overall validity of our calculations.   

Extracting the flux of 2M1207B alone requires more effort, given its proximity to the much brighter primary: we must first subtract the underlying contribution of the latter.  A simple way of doing so is suggested by a contour plot of the 2M1207 point-spread function (PSF; Fig. 3a), which reveals A's PSF to be very symmetric about the $x$ axis.  We therefore reflect the coadded $J$-image about the $x$-axis, drawn through the centroid of 2M1207A located by $IRAF$ $apphot$, and subtract the reflected image from the original.  Fig. 3b shows the residual contours after subtraction: the procedure clearly removes the primary's contribution at the secondary's position with high fidelity, leaving behind a clean image of 2M1207B on a flat, relatively null background.  B's counts are extracted from this residual image after subtracting the remaining flat background, using $apphot$; the errors are dominated by photon noise.  Relative photometry against $2M1207A$ yields $\Delta$$J$ = 7.0$\pm$0.2; adopting the 2MASS value of $J$ = 13.00$\pm$0.03 for the latter then implies $J$ = 20.0$\pm$0.2 for 2M1207B.  As a consistency check, the flux for 2M1207B was independently extracted for us by F. Marchis using the ADONIS deconvolution software; the cleaned image from this procedure was very similar to ours, and the resulting $J$-band photometry fully consistent with our value.  

Note that our relative photometry implicitly assumes that the 2MASS and VLT filters are identical (as also assumed by Chauvin et al. (2004) for their $H$ and $K_s$ photometry of this system).  Stephens \& Leggett (2004) have shown that the infrared magnitudes of L and T dwarfs are very sensitive to the precise filter bandpass used; since 2M1207B has been previously classified as an L-type, we have compared the VLT and 2MASS filters to check for any potential systematics in our photometry.  We find that the $JHK_s$ filters are very similar in the two systems: errors due to bandpass differences should be of order $\lesssim$0.1 mag, smaller than the $\pm$0.2 mag uncertainty we already quote for our relative photometry.

\subsection{Spectroscopy}
The 2-D spectra within each spectroscopic observing block (OB) were flat fielded, background-subtracted by differencing each pair of nodded images, registered and median-combined.  A 1-D spectrum of 2M1207A for each OB was then extracted directly from this final median frame, using the $apall$ task in $IRAF$.  

For 2M1207B, the proximity of the primary demands a more sophisticated extraction procedure.  While the PSF of the primary is very symmetric about the dispersion axis, self-subtraction after reflection about this axis (analogous to the technique described above for the $J$-band photometry; also employed by Close et al. 2005 for spectral extraction of AB Dor C) is not ideal for our purposes: the resulting 2M1207B spectrum is noisier than we would like in each wavelength-bin\footnote{The noise in the final residual image is always higher, in this procedure, than in the original data: the original noise at every pixel is added in quadrature to the noise at the symmetric pixel in the reflected image during subtraction.  In our extraction of the secondary's {\it integrated} $J$-band flux for photometry, this increase in noise was not a serious concern; however, it is a significant effect for {\it spectral} extraction, where we want the flux in individual narrow wavelength bins.}.  Instead, we model the PSF of the primary in detail, as described below.  

As a first step towards increasing the S/N, we smooth the median 2-D spectrum in each OB by a 5-pixel boxcar in the dispersion ($y$) direction; this correspondingly reduces our spectral resolution to $\sim$100.  The spatial PSF of 2M1207A in each row (i.e., at each wavelength) of this smoothed image is then fit by a sum of three Gaussians, plus a 2$^{nd}$ order polynomial (to account for any large-scale spatial background from chip QE effects), through a chi-squared minimization routine in $IDL$.  During fitting, the pixels corresponding to the {\it secondary's} spectrum are assigned a weight of zero, so they do not contribute to the fit.  The resulting image is our model 2-D spectrum for 2M1207A.  This is subtracted from the original smoothed image; the residual is a smoothed 2-D spectrum of 2M1207B alone.  The 1-D spectrum of 2M1207B is then extracted from this residual image with $IRAF$ $apall$.  

Fig 4 illustrates the validity of this procedure.  We display the original smoothed median 2-D spectral image for a representative OB (April 4 2005), showing the combined spectra of 2M1207A and B; our model fit to 2M1207A for this spectrum; and the residual spectrum of 2M1207B after model subtraction.  The fit is very good in the vicinity of the secondary, and the subtraction produces a relatively clean secondary spectrum on a flat null background (apart from noise, non-zero residuals remain only near the peak of the primary, far from the secondary's position).  As a second check on our extraction, we have compared the 2M1207B spectra obtained via this PSF-fitting method to the spectra derived from the simple reflected-image-subtraction method mentioned earlier.  While the fitting procedure produces much better S/N spectra, as expected, the general shape of the final spectrum is the same for both techniques, confirming the validity of our detailed fitting analysis.  

The final result of all the above steps is a set of 4 spectra each for 2M1207A and B: one for each of our 4 spectroscopic OBs.  The wavelength-scale for each spectrum is derived from Argon arc spectra.  As mentioned earlier, the 2M1207A spectra have a resolution of $\sim$550, while the 2M1207B spectra have been smoothed during extraction to $R$$\sim$100.  These spectra are dominated by telluric absorption bands, and by the wavelength-dependent spectrograph response; these effects are removed with the help of telluric standards as follows.   

Telluric standards were observed during each OB: HIP 073074 (G0V; OB on April 4 2005), HIP 059795 (G3V; OB on April 30 2005), and HIP 088409 (B3III; both OBs on June 3 2005).  Two spectra were extracted, using $apall$, for each standard: one unsmoothed for correcting 2M1207A, and one smoothed by a 5-pix boxcar, before extraction, for correcting 2M1207B. HIP 088409 is moreover slightly extincted, by $\av$$\approx$1.0 (revealed by our comparison to the colors of unreddened B3 stars, and also cited by Rubin et al. 1962), producing a small change in the $HK$ spectral slope; we have dereddened its spectrum accordingly. Next, narrow intrinsic spectral lines (e.g., Br$\gamma$ absorption) in the telluric standards must be removed, to avoid producing spurious features in the science data upon telluric-correction.  We accomplish this by dividing our telluric standard spectra by normalized (continuum divided out) spectra of appropriate spectral-type standards, rebinned to our resolution.  For HIP 073074 and 059795, our spectral standard is a solar-spectrum (G2V; NSO/Kitt Peak FTS data, produced by NSF/NOAO and available on the VLT site); for HIP 088409, we use a spectrum of HD 35653 (B0.5V) from the near-IR spectral library of Lancon et al. (1992).  Finally, the intrinsic spectral slope of the telluric standards is removed by dividing by a blackbody of the appropriate \teff: 5930K, 5785K and 17000K for HIP 073074, 059795 and 088409 respectively.  This leaves behind pure telluric absorption spectra, convolved with the wavelength-dependent spectrograph response.  The 2M1207A and B spectra are divided by these to produce the final telluric- and spectrograph-response-corrected spectra for each OB.  

The resulting $HK$ spectra of 2M1207A from 3 of our 4 OBs are nearly identical, and also agree very well with an $HK$ spectrum of this source recently obtained with CORMASS on Magellan (K. Luhman, pvt. comm., 2005).  The spectrum from the remaining OB (April 4, 2005), however, shows a distinct linear change in slope compared to the other three.  The spectrum of the secondary from the same OB also evinces an analogous slope-offset relative to the others.  Such an effect is expected from imperfect centering of the source in the narrow slit.  Given the close agreement between three of our 2M1207A spectra and the independent CORMASS result, it is safe to assume that only the April 4 data are anomalous.  A linear slope correction factor is therefore derived for the April 4 2M1207A spectrum, by ratioing this spectrum with the median of the other three and fitting a straight line to the result.  This correction factor is then applied to the April 4 spectra of both 2M1207A and B.  

Lastly, the spectra from the 4 OBs are normalized by their means over 2.1--2.2$\mu$m and median-combined, producing one final spectrum each for 2M1207A and B; these are the spectra used in our analysis.  The S/N is $\sim$30 for the primary and $\sim$3--10 for the secondary.  

\subsection{Astrometry}
Our observations were designed for photometry and spectroscopy (the main topic of this paper) and not astrometry.  Consequently, while the VLT provides nominal plate-scale and plate-rotation values for the CONICA S27 camera used for our $J$-band imaging, we have not obtained reference frames to estimate the errors in these quantities.  Nevertheless, comparing them to the values found by Chauvin et al. (2004), from reference images with the same camera, lets us set conservative error bars as follows, and still perform useful astrometry.

The nominal plate-scale and plate-rotation for CONICA S27 are 27.150 mas/pix and 0$^{\circ}$ (North up, East left) respectively.  For the same camera, Chauvin et al. (2005a) cite a plate-scale of 27.010$\pm$0.05 mas/pix over their 1 year of observations, and a counter-clockwise plate-rotation of 0.07$\pm$0.10$^{\circ}$ to 0.14$^{\circ}$$\pm$0.10$^{\circ}$.  For our calculations, therefore, we use the nominal S27 values, but conservatively adopt error bars equal to the maximum difference between Chauvin et al.'s estimates for these parameters (including their error bars) and the nominal ones.  That is, we use plate-scale = 27.150$\pm$0.20 mas/pix and rotation = 0$^{\circ}$$\pm$0.25$^{\circ}$.  

\section{Models Employed: Synthetic Spectra and Evolutionary Tracks}
We compare the observed spectra to the latest synthetic spectra generated by the Lyon group with the PHOENIX code (Allard et al. 2001).  Specifically, we use the models designated DUSTY-2000, COND-2002 and SETTL-2005 (Allard et al. 2001; Allard et al. 2003).  These incorporate the most recent AMES line-lists for both TiO (Langhoff 1997, Schwenke 1998) and \h2o (Partridge \& Schwenke 1997).  TiO opacity dominates in the optical, while \h2o dominates in the infrared.  The models include about 500 million molecular lines ($\sim$307 million of \h2o and $\sim$172 million of TiO), the formation of over 600 gas-phase species and 1000 liquids and crystals, and opacities of 30 different types of grains (Allard et al. 2000a; Allard et al. 2001; though some important species remain absent: see \S7.1.3).  All models used here are for solar-metallicity ([M/H] = 0.0).  

Convection in these models is handled with mixing-length theory (MLT), which seems to reasonably approximate the true convective transport mechanism (Chabrier \& Baraffe 2000).  The DUSTY-2000 models use $\alpha$ = 1.0 (where $\alpha$ $\equiv$ $l$/$H_P$, with $l$ being the mixing-length and $H_P$ the pressure scale height).  However, recent full 3-D radiative hydrodynamical simulations indicate that $\alpha$ $\approx$ 2 is a better approximation for mid-M types, both in the field as well as in the PMS phase (Ludwig 2003).  Such simulations offer the best insight so far into the actual convection process (see Chabrier \& Baraffe 2000); consequently, the COND-2002 and SETTL-2005 models, which are more recent than DUSTY-2000, use $\alpha$ = 2.  In this context, we note that our TWA sources are expected to have an age $\sim$ 5--10 Myrs, and thus (according to theoretical evolutionary tracks) log $g$ $\approx$ 4$\pm$0.25; the secondary is also expected to be very cool ($<$ 2000K) from the previous spectral type analysis by Chauvin et al. (2004).  Baraffe et al. (2002) show that the value of $\alpha$ affects H$_2$ formation, and thus the evolutionary path (e.g., \teff, \lbol), for extremely young ($\sim$ 1 Myr-old), very low-gravity (log $g$ $\leq$ 3.5) objects with \teff $\sim$ 2200--4000K.  However, they also show that the effect of $\alpha$ on evolution is negligible both by an age of $\sim$10 Myr (log $g$ $>$ 3.5) and for \teff $<$ 2200K: at these higher gravities and/or lower \teff, H$_2$ forms efficiently regardless of whether $\alpha$ is 1 or 2.  Thus for both our sources, and especially for the secondary, the choice of $\alpha$ is not crucial.  Consistent with this, we will explicitly show that both DUSTY-2000 and SETTL-2005 spectra imply very similar \teff for this object, in spite of the difference in $\alpha$ between the two sets of models.

The fundamental distinction between the DUSTY, COND and SETTL synthetic spectra lies in their treatment of photospheric dust.  All three models treat grain formation self-consistently, through chemical equilibrium calculations, and thereby also account for the depletion of chemical species that become sequestered in grains (Allard et al. 2001).  However, the DUSTY models assume that all the grains that form remain suspended in the photosphere; thus, these models include dust opacity as well.  Conversely, the COND models assume that all the grains have gravitationally settled below the photosphere, and therefore neglect grain opacity altogether.  These two scenarios represent the two extremes of dust behaviour expected in cool dwarfs, as photospheric grains first form and then, with decreasing temperature, gradually drain out of the atmosphere.  Finally, the SETTL models attempt to capture the true behaviour of grains more faithfully, by self-consistently including the settling process itself, in addition to dust opacity.  Thus, at relatively warm \teff, when grains have formed and remain predominantly suspended in the atmosphere, or at very cool \teff, when grains have completely rained out of the photosphere, the SETTL spectra resemble DUSTY and COND respectively; in the intervening temperature regime, when grains have partially settled, SETTL spectra are intermediate between these two extreme models.  Note that for \teff $\gtrsim$ 2500K, the chemical equilibrium calculations indicate that {\it no} dust forms at all, so DUSTY, SETTL and COND models are all exactly the same at these higher \teff.  

The treatment of dust settling in the SETTL models is based on the 3-D radiative hydrodynamic (RHD) simulations by Ludwig (2003) and Ludwig et al. (2006); a full description can be found in the latter two articles and in Allard et al. (2003).  In brief, the mixing of atmospheric material, including grains, by convective overshoot motions can be parametrized by $v_{mix}$($z$) $\equiv$ $F_{mass, up}$($z$)/$<{\rho}{\rm{(}}z{\rm{)}}>$.  Here $v_{mix}$ is the velocity of mixing motions at some atmospheric height $z$, $F_{mass, up}$($z$) is the upward-directed mass flux at that location, and $<{\rho}{\rm{(}}z{\rm{)}}>$ is the average mass density there.  Note that more mixing is equivalent to less settling.  The RHD simulations show that $v_{mix}$ declines exponentially with atmospheric height above the Schwarzschild (convective-stability) boundary.  To a good approximation, the length-scale of this exponential decline is found to be $H_{mix}$($z$) $\propto$ $H_P$($z$)$\sqrt{g_0/g}$, where $H_P$($z$) is the pressure scale height at $z$, and $g$ is the surface gravity in c.g.s. units, normalized to a gravity of $g_0$ = 10$^5$ cm s$^{-2}$ (i.e., log $g_0$ = 5).  Thus the simulations indicate that, all else being equal, there is greater mixing ({\it less} settling) at lower gravities.  In the M dwarf simulations by Ludwig (2003), the constant of proportionality in the above relationship is $\sim$0.5, i.e., $H_{mix}$($z$) $\approx$ 0.5$H_P$($z$) for log $g$ = 5. 

We have used two sets of SETTL-2005 spectra based on this formulation.  In the first, mixing is parametrized by $H_{mix}$($z$) = 0.5$H_P$($z$)$\sqrt{g_0/g}$, with log $g_0$ = 5.  As described above, this is the result directly implied by the RHD simulations.  In the second, $H_{mix}$($z$) = 0.25$H_P$($z$)$\sqrt{g_0/g}$.  That is, the mixing scale-height adopted in the latter models for the convective overshoot is half that indicated by the RHD calculations, thereby simulating less mixing ({\it more} settling).  We note that the SETTL-2005 models we have described and used here differ slightly, in the settling parametrization employed, from the SETTL-2002 ones on the Lyon group's website ({\it ftp://ftp.ens-lyon.fr/pub/users/CRAL/fallard}).  The 2005 models were kindly supplied to us by D. Homeier and F. Allard (pvt. comm. 2006).  

An alternate prescription for describing grain settling has been put forward by Ackerman \& Marley (2001), who parametrize the process through a grain sedimentation efficiency factor, $f_{sed}$.  For the sake of completeness, we have compared our observations to these synthetic spectra as well, kindly supplied to us by D. Saumon (pvt. comm. 2006).

Finally we also compare the data to Lyon theoretical evolutionary tracks, based on both DUSTY and COND atmospheres (Chabrier et al. 2000; Baraffe et al. 2003).  In these evolutionary models, the atmospheric opacity acts as a boundary condition on the interior calculations; thus, for a given mass, the temporal evolution of the {\it spectral} and {\it photometric properties} does depend on the flavor of atmosphere adopted (DUSTY or COND).  However, the {\it cooling properties} of brown dwarfs -- i.e., the temporal evolution of \teff and \lbol, and thus of radius and surface gravity -- are negligibly affected by the opacity: \teff($t$) $\propto$ ${\kappa}_R$$^{{\sim}1/10}$ and \lbol($t$) $\propto$ ${\kappa}_R$$^{{\sim}1/3}$ (Burrows \& Liebert 1993; Baraffe et al. 2002).  Consequently, we will distinguish between the evolutionary tracks based on COND and DUSTY when comparing to the observed photometry of our sources, but not when comparing to their \lbol or \teff.

We note that the $JHK_s$ photometry for 2M1207A and B, as well as for AB Pic B which we also analyse, have been obtained in the 2MASS system, while the Chabrier et al. (2000) and Baraffe et al. (2003) evolutionary models supply predicted $JHK$ in the CIT system.  We have therefore converted the 2MASS values for 2M1207A, a late M type source, to CIT using the transformations given by Carpenter (2001).  For the L-type sources 2M1207B and AB Pic B, we have converted from 2MASS to CIT using the transformations given by Stephens \& Leggett (2004).  The resulting shift from 2MASS to CIT is $\sim$0.1 mag in $J$ and $<$0.05 mag in $H$ and $K$ for all three objects; for 2M1207B and AB Pic B, these shifts are at most comparable to, or much less than, the uncertainties in their photometry.

The transformations derived by Stephens \& Leggett (2004) are dependent on spectral type.  For the conversions above, we have therefore nominally assumed L7.5 for 2M1207B and L1 for AB Pic B, to be consistent with the $\sim$L5--L9.5 and $\sim$L1${^{+2}_{-1}}$ types assigned to them by Chauvin et al. (2004) and Chauvin et al. (2005b) respectively.  In reality, the transformations are quite flat in the L regime: varying the assigned type over the entire L range changes the corresponding 2MASS-to-CIT transformations by $\lesssim$0.05 mag (see Fig.6 of Stephens \& Leggett).  This is much less than the photometric error bars for 2M1207B and AB Pic B, so the uncertainty in their spectral type does not affect our analysis.  Of more concern is the fact that the Stephens \& Leggett transformations are derived for higher gravity field L and T dwarfs.  Our objects are much younger and lower gravity ones, with correspondingly somewhat different spectral shapes (as discussed later), and hence presumably not subject to exactly the same filter transformations as the field dwarfs.  Thus ideally we should compare our data to evolutionary models with predicted photometry in the same filter system as the observations (2MASS), instead of employing the reverse and more uncertain operation of transforming our data to the models' filters (CIT).  While time constraints prevented us from obtaining the entire suite of evolutionary models in 2MASS filters, therefore, we checked for potential systematics by computing 2MASS and CIT photometry for low and high gravity synthetic spectra -- the same COND and DUSTY spectra used by the evolutionary models for photometry predictions --  sampling late-M to mid-T temperatures ($\sim$2500--1000K).  We found that while gravity does make a small difference, it is negligible for our purposes: the shifts from 2MASS to CIT magnitudes for the low gravity (log g = 4.0) synthetic spectra are in the same direction as, and within 0.05--0.1 mag of, the shifts predicted by the Stephens \& Leggett transformations for higher gravity (log g = 5--5.5) field objects in the same spectral type range.  These variations are again smaller than the uncertainties in the photometry of 2M1207B and AB Pic B; consequently, we are justified in simply adopting the Stephens \& Leggett transformations for the purposes of this paper.

For 2M1207AB, we also show ground-based photometry in $L'$ (Johnson-Glass system) and HST photometry in the F090M and F160W filters.  We do have Lyon tracks with predicted photometry in these filters, so no data transformations were required for these.   

Lastly, we note that all magnitudes cited are relative to Vega.

\section{Results}

\subsection{Common Proper Motion and $J$-band Photometry}
Using 3 VLT astrometric observations from April 2004 to March 2005, Chauvin et al. (2005a) found that 2M1207AB very likely comprises a bound co-moving system.  With an improved estimate of the distance, and 2 additional HST observations in August 2004 and April 2005, Song et al. (2006) have confirmed this result to 16$\sigma$.  Our $J$-band image from March 26 2005 provides an additional data point in this series.  We find a separation of 769$\pm$10 mas between 2M1207A and B (including centroiding errors), and a position angle of B with respect to A of 125.6$^{\circ}$$\pm$0.7$^{\circ}$. These are completely consistent with Song et al.'s values, from the combined VLT and HST data, of sep = 773.0$\pm$1.4 mas and PA = 125.37$^{\circ}$$\pm$0.03$^{\circ}$.

Our relative photometry between 2M1207A and B implies ${\Delta}J$ = 7.0$\pm$0.2; with $J$ = 13.00$\pm$0.03 from 2MASS for the primary, we have $J$ = 20.0$\pm$0.2 for 2M1207B.  For comparison, Song et al.'s (2006) HST NICMOS photometry with the F110M filter (the closest approximation to $J$ in their observations) implies $\Delta$F110M = 7.17$\pm$0.15, and F110M = 20.61$\pm$0.15 for 2M1207B.  These results are consistent with ours: the F110M filter (passband $\sim$1.0--1.2$\mu$m) covers only the blueward side of the $J$-band, while both the F110M and $J$ magnitudes we cite are relative to the A-type star Vega, which is very blue across the $J$-band; the result is a fainter F110M magnitude for both objects compared to $J$. 

\subsection{Spectral Types}
\subsubsection{Previous Results}
Gizis (2002) assigned a spectral type of M8 to 2M1207A, based on a comparison of various atomic and molecular indices in its low-resolution optical spectrum to those of field M dwarfs.  This procedure is not ideal, since the optical spectra of very young, low-gravity late-M types are known to diverge somewhat from those of higher-gravity field dwarfs.  Optical types for such young objects have therefore also been obtained by fitting templates composed of combined M-type giant and dwarf spectra (Luhman 1999; Luhman et al. 2003; White \& Basri 2003).  Nevertheless, the resulting error in Gizis' typing appears small: as he shows, the gravity-sensitive VO indices in 2M1207A's optical spectrum, while stronger than in field M dwarfs (as expected), are comparable to those in the very young brown dwarf GY141, which Luhman et al. (1997) have assigned a type of M8.5 through dwarf+giant fitting.  Similarly, Mohanty et al. (2005) report that the temperature-sensitive TiO bands in high-resolution optical spectra of 2M1207A are well-matched by those in M8$\pm$0.5 young brown dwarfs, where the latter have again been typed previously through dwarf+giant fitting.  A spectral type of M8$\pm$0.5 thus appears acceptable for 2M1207A.  

The situation has been more uncertain for 2M1207B.  In the discovery paper, Chauvin et al. (2004) showed that its $H$-$K$ and $K$-$L$ colors were comparable to, or redder than, those of the reddest late-L field dwarfs.  Meanwhile, their spectrum for it in the $H$-band alone, at low S/N, appeared similar to that of L5--L9.5 field dwarfs.  The colors implied by Song et al.'s (2006) HST NICMOS photometry, covering roughly $I$ to $H$ bands, appear consistent with this mid-to-late L classification.

\subsubsection{New Infrared Spectra}
Fig. 5 shows our $HK$ spectra of 2M1207A and B compared to those of field M6--L8 dwarfs.  For 2M1207A, we see that the general slope of the spectral energy distribution (SED) from $H$ to $K$, the shape of the $K$-band spectrum, and the depth of the CO bandhead at 2.3$\mu$m are all comparable to those in the M8 dwarf (note that the CO bandheads at longer wavelengths are degraded by noise in 2M1207A).  We emphasize that an exact determination of infrared spectral type is not our goal here; we are simply pointing out that the overall appearance of the $HK$ spectrum of 2M1207A is reasonably similar to that of an M8 dwarf, consistent with the optical spectral type assigned earlier.  

However, there are also notable departures from the dwarf spectra.  The most striking of these is the strongly peaked $H$-band profile in 2M1207A, in sharp contrast to the more rounded profiles in all the other dwarfs.  Such a triangular $H$-band shape seems to be a distinguishing feature of young late-type objects; for instance, it has been noted earlier in very young and cool brown dwarf candidates in the Trapezium cluster (Lucas et al. 2001).  As Kirkpatrick et al. (2006) point out, this effect is related to the decrease in H$_2$ collisionally-induced absorption (CIA) opacity at low gravities (\S7.1.1; see also Borysow et al. 1997).  Another marker of low-gravity in 2M1207A is the very weak NaI absorption at 2.2$\mu$m, compared to the stronger absorption in the higher gravity field dwarfs.  
      
The spectrum of 2M1207B is even more arresting.  The deep \h2o absorption troughs in the $H$-band are commensurate with a mid-to-late L type object, as suggested by Chauvin et al. (2004); so is the large \h2o opacity in the $K$-band.  However, despite the relatively low S/N (3--10$\sigma$), a triangular $H$-band profile is clearly visible, as in the primary and unlike any of the M or L dwarfs, indicating low gravity.  Furthermore, 2M1207B exhibits a strikingly different $K$-band morphology compared to the field L dwarfs.  Its spectral energy distribution in $K$ peaks at a significantly longer wavelength, and its $K$-flux relative to $H$ is much higher, than in the field dwarfs.  The $H$-$K$ color of 2M1207B is consequently far redder than in typical L dwarfs, explaining the results of Chauvin et al.'s (2004) photometry.  These $K$-band effects arise from a combination of increasing photospheric dust and changes in H$_2$-CIA and \h2o opacity with decreasing gravity (\S7.1.1).  Notice that the CO bandhead at 2.3$\mu$m also appears much stronger in 2M1207B than in the L dwarfs (although we caution that poor S/N in this region of the secondary's spectrum may also be a contributing factor).  In M types, deeper CO is associated with low gravity (e.g., Luhman \& Rieke 1998); if this trend continues into the L class (as the atmospheric models suggest it does; \S7.1.1), it would be yet another signature of low gravity in 2M1207B.  

In this context, it is noteworthy that anomalously red IR spectra and deep CO have also been seen in the $\sim$L4.5 field object 2MASS 2224-0158 (Cushing et al. 2005); similarly red colors and a triangular $H$-band identified in the $\sim$L0 source 2MASS 0141-4633 (Kirkpatrick et al. 2006); and anomalously red colors, a triangular $H$-band and deep CO seen in the L6.5 field dwarf 2MASS 2244+20 (McLean et al. 2003).  Substantial photospheric dust and low gravity have been proposed to explain the first, low gravity to explain the second, and both effects may be at play in the third, in step with our conclusions here.  

To summarize: while the general strength of \h2o opacity in its $H$ and $K$ bands indicates a mid-to-late L type for 2M1207B, the shape of its $H$-band and the overall form of its SED from $H$ to $K$ depart strongly from typical L dwarf spectra.  As such, {\it it is impossible to assign a precise spectral type to 2M1207B based on $HK$ comparisons to field L dwarfs}.  It is more fruitful to investigate the physical properties of 2M1207B (and A) directly, through comparisons to synthetic spectra and evolutionary models.

\section{Analysis: Effective Temperature and Mass}

\subsection{Effective Temperature from HK Spectra}

\subsubsection{Trends in the Synthetic Spectra}
To supply a physical basis for interpeting synthetic spectral fits to our data, we first discuss some salient trends in the model spectra with varying \teff, gravity and dust.  Fig. 6 shows DUSTY and COND models (described in \S5) for various \teff (1600, 2000, 2400K) and two different gravities (log $g$ = 4.0, 5.5).  The following behaviors can be seen.  

\noindent{\it (1)} Dust starts to form in DUSTY and COND models only at \teff$\lesssim$2500K.  At higher temperatures, therefore, the two models are identical (not shown); at \teff below but still close to 2500K, they remain very similar since little dust has formed yet (e.g., 2400K shown).

\noindent{\it (2)} At a fixed gravity, DUSTY models become redder with decreasing \teff: the combination of large dust opacity at shorter wavelengths, and suppressed \h2o opacity at longer wavelengths in the presence of dust (backwarming by grains both dissociates \h2o and creates a hotter, more transparent \h2o opacity profile) pushes the emergent flux redwards.  The effect is exacerbated at lower \teff as more grains form and remain suspended in the photosphere.  Conversely, COND models become bluer in the near-IR with decreasing \teff: without dust opacity (grains form but settle out), increasing \h2o and H$_2$-CIA opacity at longer wavelengths forces emergent flux bluewards, and the effect grows as the two opacities strengthen with falling \teff (at sufficiently low \teff, CH$_4$ absorption further magnifies this spectral bluing in the COND models). The crucial result, for our purposes, is that DUSTY models are always redder than COND ones in $H$-$K$, at a given \teff and gravity.  

\noindent{\it (3)} In both COND and DUSTY spectra, lower gravity produces a more triangular $H$-band profile, and shifts the peak in $H$ and $K$ redwards.  This can be understood as follows. On the one hand, the overall \h2o opacity in the $H$ and $K$ bands decreases with decreasing gravity, allowing more flux to emerge in these regions.  Simultaneously, H$_2$-CIA opacity (due to induced 1-0 quadrupolar moment of H$_2$), which peaks at $\sim$2.5$\mu$m and declines at shorter wavelengths, also diminishes at lower gravities (Borysow et al. 1997).  The combination of the two produces the observed changes in the $H$ and $K$ band shapes.  For comparison, it is interesting to note that very {\it large} H$_2$-CIA (higher than in normal field L dwarfs) is implicated in the flat and suppressed $K$-band and blueward-slanted $H$-band of metal-poor L dwarfs, such as 2MASS 0532+8246 (Burgasser et al. 2003b).  These effects are opposite to those described here for low gravity / low H$_2$-CIA.  This is reasonable, since low metallicity produces higher pressures at a given optical depth, while low gravity implies lower pressures, and it is fundamentally pressure that drives the behavior of H$_2$-CIA.   

\noindent{\it (4)} Finally, the 2.3$\mu$m CO bandhead becomes deeper with lower gravity at a given \teff, in both COND and DUSTY models.  In the CO region, a major continuum opacity source is \h2o; at the lower pressures implied by lower gravity, the opacities of both molecules decrease.  However, since \h2o is a triatomic molecule compared to the diatomic CO, its opacity decreases faster than that of CO.  Consequently, the CO bandheads appear deeper at lower gravity relative to the surrounding continuum.  

\subsubsection{Model Fits to HK Spectra: Procedure}
We wish to determine \teff for 2M1207A and B from fits to synthetic spectra.  As shown above, the shape of the spectrum is affected by temperature as well as gravity; to determine both {\it simultaneously} through spectral fitting requires analysis of spectral features at higher resolution and higher signal-to-noise than we possess.  Instead, since it is temperature we are after, we a priori {\it fix} the gravity to the range implied by the evolutionary models for substellar objects at the age of 2M1207AB.  This clearly makes our results dependent on the evolutionary models from the outset.  However, since our subsequent mass estimates (as well as all analyses of this system by other researchers) are based on these models anyway, adopting them from the beginning is self-consistent, and not an additional drawback.  
   
For ages of 5--10 Myrs, the evolutionary models predict gravities of $\approx$ 4.0$\pm$0.25 (log $g$ in cgs) over the entire substellar domain.  We therefore employ synthetic spectra with log $g$ = 3.5--4.5 (step 0.25 dex) in our \teff analysis.  At each gravity, the observed $HK$ spectra are compared to a range of model \teff: 3000--2000K DUSTY spectra for 2M1207A, and 2500--1500K DUSTY spectra and 2500--900K COND spectra for 2M1207B.  The best-fit model spectrum is found via least-squares fitting to $H$ and $K$ simultaneously, as follows.  

The model spectra are rebinned to the spectral resolution of the data ($R$ $\sim$550 and $\sim$100 for 2M1207A and B respectively).  Next, as a first guess for the appropriate normalization for comparing models and data, both are initially normalized by the mean of their respective fluxes over the range 2.1--2.2$\mu$m, near the peak of the $K$-band.  However, while the current synthetic spectra are very sophisticated, they still contain significant opacity uncertainties (e.g., Allard et al. 2001; Leggett et al. 2001).  Forcing concordance a priori between the models and data over a narrow wavelength range can thus lead to skewed results.  Consequently, we allow the normalization of every model to vary from 0.5 to 2 times the initial guess, in steps of 1\%, during the fitting process.  This allows us to optimize the global fit of each model spectrum to the $HK$ data.  Our least-squares statistic is defined as: 

$$ s^2 \equiv \frac{\sum_{i=1}^{N}[(n*f^{model}_i - f^{obs}_i)/f^{obs}_i]^2}{N}  $$

where $f^{model}_i$ and $f^{obs}_i$ are the initially normalized model and observed fluxes in each wavelength bin $i$; $n$ is the varying normalization then applied to the model, ranging from 0.5 to 2 in steps of 0.01; and $N$ is the total number of wavelength bins. $s^2$ therefore captures the average fractional deviation between the data and a model for a specified model normalization.  For any given \teff/log$g$ model, the optimum fit to the data is obtained by finding the normalization $n$ which minimizes $s^2$.  We note that the optimum $n$ generally differs from our initial guess by $<$20\%.  The overall best-fit \teff/log$g$ model is then identified as the one with the minimum least-squares value $s^2$ among all these normalization-optimized model spectra.  

During the fitting, we mask regions of high telluric absorption (1.82--1.98$\mu$m) and very low S/N ($>$ 2.5$\mu$m for 2M1207A, $>$2.25$\mu$m for 2M1207B, and $<$1.41$\mu$m for both).  Moreover, the model fits to 2M1207A are problematic over the range 1.5--1.7$\mu$m, as discussed below; we have thus masked this region as well for the 2M1207A fits.  
 
Finally, the validity of the best-fit obtained, and the range of admissible \teff values, is verified by examining all the fits by eye.  

\subsubsection{Model Fits to HK Spectra: Results}

{\it Primary}: For 2M1207A, Fig. 7a shows the least-squares contours for normalization-optimized DUSTY models, with \teff = 2000--3000K and log $g$ = 3.5--4.5.  The important quantity here is not the absolute, but the relative value of the least squares, which signifies the relative merit of the fits; to portray this clearly, all the least-squares values $s^2$ have been normalized (divided) by the global minimum in $s^2$, obtained for the best-fit model.  The best fit is nominally at \teff = 2650K, log $g$ = 4.0; however, we see that the fits are rather insensitive to gravity, and \teff $\approx$ 2650K is the best match from log $g$ = 3.5--4.5.  Changing the temperature by $\pm$200K from this value worsens the fit significantly. 
 
These conclusions are verified by examining the fits by eye.  Fig. 8 shows the model comparisons to the data at 2650$\pm$200K and log $g$ = 3.5, 4.0 and 4.5.  At the best-fit \teff of 2650K, the models match the $K$-band flux, the redward side of the $H$-band, and the \h2o absorption feature at $\leq$1.5$\mu$m rather well at these gravities.  The 2450K models overestimate the \h2o opacity at $\leq$1.5$\mu$m, and begin to deviate from the observed shape of the $K$-band, while the 2850K models underestimate the \h2o opacity as well as the $K$-band flux.  From these fits, we would be justifed in adopting \teff $\approx$ 2650$\pm$200K.  

However, notice that all the models overestimate the 1.5--1.7$\mu$m flux, near the peak of the $H$-band (as noted above, this region is excluded in the least-squares fitting).  The \teff = 2650K, log $g$=3.5 model comes closest, but is still not an adequate match.  Such over-prediction of $H$-band emission by the model spectra has been pointed out before, in comparisons to both field and young M-types (e.g., Leggett et al. 2001; Lucas et al. 2001), and is ascribed to known inadequacies and incompleteness in the current \h2o linelists in the models at these relatively hot temperatures (Allard et al. 2001).  Moreover, a forest of FeH lines dominating the $\sim$1.58--1.75$\mu$m region in field M and L dwarfs has recently been identified (Wallace \& Hinkle 2001; Cushing et al. 2003).  These lines (thought to arise from the 0--0 band of the $E^4\Pi$--$A^4\Pi$ system) are not included in the current generation of synthetic spectra at all, and would also tend to depress the model flux precisely where it is anomalously high now.

Given these \h2o and FeH opacity uncertainties in the infrared, one would ideally like to verify our $HK$-based \teff with a more reliable diagnostic.  TiO bandheads in the optical provide just such a tool.  Allard et al.(2000) showed that the new AMES-TiO linelists in the synthetic spectra provide a much better fit to the optical SEDs and photometry of M dwarfs.  Mohanty et al. (2004a) subsequently demonstrated that the models match quite well the observed TiO bandheads at $\sim$8400\AA, in {\it high}-resolution spectra of very young mid-to-late M sources.  Moreover, they found that this TiO-band (specifically, triple-headed band at $\lambda\lambda\lambda$8432, 8442, 8452\AA, identified as the $E^3\Pi$--$X^3\Delta$ system by Solf 1978) is highly temperature-sensitive and simultaneously relatively gravity-{\it in}sensitive, ideal for fixing \teff.
   
We have previously obtained high-resolution optical spectra of 2M1207A on Magellan (Mohanty et al. 2003).  Following Mohanty et al. (2004a), therefore, we derive \teff by comparing its TiO bandheads to synthetic spectra.  The results are plotted in Fig. 9.  The best-fit model, at 2500K, clearly matches the data remarkably well: both in the TiO bandheads (except a small mismatch to the very core of the $\lambda$8432 bandhead\footnote{The 2500K model is slightly deeper than the data in  the core of the $\lambda$8432 bandhead, which appears weakly gravity-sensitive over the 1 dex range in log $g$ plotted: while the fit to the core is best at log $g$ $<$ 4.5 (consistent with the evolutionary model prediction of log $g$$\sim$4.0 for brown dwarfs at 5--10 Myrs), a slight mismatch persists down to log $g$=3.5 (the lowest gravity we examine).  The 2500K model also does not completely reproduce the relatively weak absorption feature at 8440\AA, just blueward of the $\lambda$8442 bandhead.  However, such small deviations between the data and model spectra are to be expected, even for the best fits, considering the huge number of TiO line opacities involved in matching the data at high spectral resolution.}) and in the surrounding continuum.  Changing \teff by only $\pm$50K produces small but noticeable departures from the data, in the $\lambda\lambda$8442, 8452 band strengths as well as in the average continuum flux redward of $\lambda$8452, while $\pm$100K changes yield clearly worse fits. 

From the TiO data alone, therefore, we would infer \teff = 2500$\pm$100K for 2M1207A.  Crucially, this is within 150K of the 2650K derived from the $HK$ low-resolution data, and well within the combined errors of the optical and infrared fits (100--200K in each).  This confirms the general validity of the $HK$ fits (at the $\sim$200K precision we desire in this work) despite the remaining uncertainties in the model \h2o and FeH opacities.  

From the above analyses, the most conservative estimate of 2M1207A's temperature would be \teff = 2400--2850K, the union of the full ranges implied by the optical and NIR fits.  The lower limit, compatible with both the TiO and $HK$ data, is acceptable.  The upper limit though, which comes from the $HK$ fits alone, appears too high: On the one hand, the $HK$ fits become less sensitive to \teff above $\sim$2700K (as shown by the contours in Fig.7a, which become much more widely spaced at higher \teff than at lower ones); on the other hand, the TiO fits, which do remain very sensitive to higher temperatures, appear incompatible with \teff much higher than $\sim$2600K (as shown by the large mismatch between the models and TiO data at 2700K, in Fig. 9).  A safe but better estimate of the \teff range is therefore $\sim$2400--2700K, which includes the best-fit \teff implied by both the optical and NIR, and the best limits compatible with both.  Thus we finally adopt \teff $\approx$ 2550$\pm$150K for 2M1207A (where the mean is the middle of the range, and also falls between the TiO and $HK$ best-fits).  As we will demonstrate later, this estimate is also in agreement with the NIR colors of 2M1207A ranging from $J$ to $L'$ (\S7.4).  

{\it Secondary}:  For 2M1207B, Fig. 7b show the least-squares contours for normalization-optimized DUSTY models with \teff = 1500--2000K and log $g$ = 3.5--4.5. Again, all the least-squares values $s^2$ have been normalized by the global minimum in $s^2$, obtained for the best-fit model.  As in the primary, we see that the fits are rather insensitive to gravity over the range considered; the best fit to the DUSTY models is at \teff $\approx$ 1600K from log $g$ = 3.5--4.0, rising to $\sim$1650K at log $g$ = 4.5.  Changing the temperature by $\pm$100K from this value worsens the DUSTY fits significantly.  We note that all the COND fits are far worse, and not shown (but see below and Fig. 10).  

These conclusions are verified by examining the fits by eye.  Fig. 10 shows the model comparisons to the data for \teff = 1500--1800K and log $g$ = 3.5, 4.0 and 4.5.  Despite the noise in the data, it is evident that the DUSTY 1600K models reproduce the general shape of the $HK$ spectrum very well at these gravities; the 1500K DUSTY models are clearly redder, and the 1700K models clearly bluer, than the data.  Since DUSTY models become bluer with increasing temperature, \teff significantly higher than 1700K is inadmissible (as shown in Fig. 10 by the very poor match to 1800K DUSTY).  Similarly, since the 1500K model is already too red, and DUSTY spectra become redder with decreasing \teff, temperatures significantly lower than 1500K are not admissible either, in the context of DUSTY models.  Conversely, the 1500K COND model is far bluer than the data, and also a poor match to the individual $H$ and $K$-band profiles; since these models become even bluer with lower temperature, \teff $<$ 1500K yields even poorer COND fits (not shown).  

In Fig. 11, we further compare our data for 2M1207B to two flavors of SETTL models (differing by a factor of 2 in the assumed convective mixing efficiency in the overshoot layers; see \S5), over a range of \teff at log $g$ = 4.0.  In both cases, we see that models with \teff $\approx$ 1500--1700K provide acceptable fits to the data, while cooler ones deviate sharply.  In particular, the strong CH$_4$ absorption that depresses the flux at $\geq$2.2$\mu$m in the synthetic spectra for \teff $<$ 1500K is notably absent in 2M1207B.  We also note that the more efficient mixing (less settling) model matches the spectral shape slightly better, consistent with our earlier finding that DUSTY ({\it no} settling) spectra provide good fits to the data at the same \teff.  Very similar results are obtained on comparing our data to the $f_{sed}$ models (kindly provided to us by D. Saumon; not plotted): \teff $<$ 1500K is not compatible with the data, and good fits are obtained with significantly less sedimentation ($f_{sed}$ = 2) than required for field L and T dwarfs ($f_{sed}$ $\sim$ 3--5; Marley et al. 2002, Knapp et al. 2004).    

Putting all this together, we conclude that \teff $\approx$ 1600$\pm$100K is appropriate for 2M1207B, and that 1500K is a lower bound on plausible \teff for this object regardless of which spectral models are used.  Moreover, the model comparisons strongly indicate a dusty atmosphere with little grain sedimentation.  The diminished degree of settling, compared to field L dwarfs at similar \teff, is not very surprising, since the surface gravity of 2M1207B (based on the evolutionary models for its age) is lower by at least an order of magnitude.  Henceforth, therefore, we refer only to the full DUSTY models.  

As an aside, we note that the $\sim$1600K DUSTY and SETTL models that best fit 2M1207B's $HK$ spectrum are able to fit the peak of the $H$-band adequately: there is no obvious sign that the synthetic spectra at these temperatures overestimate the flux over 1.5--1.7$\mu$m, unlike in the case of 2M1207A at higher \teff.  Consequently, we have not masked out this region during our model-fitting to 2M1207B (but see below).  This lack of any obvious problem in the secondary, compared to the primary, is possibly connected to the behaviour of FeH opacities: in field dwarfs, FeH bands are very strong in the $z$-, $J$- and $H$-band spectra of late M dwarfs, but weaken considerably by mid-to-late L (Burgasser et al. 2002; Cushing et al. 2003).  If this trend holds for young M and L dwarfs as well, it could explain our result: since the $H$-band FeH opacities are not yet incorporated in the current synthetic spectra, a mismatch would be seen for the late M primary where these opacities are important, but not in the mid-to-late L secondary where the FeH absorption is much weaker.  

On the other hand, one might postulate that a problem with the predicted flux at the $H$-band peak does persist for the secondary as well.  That is, perhaps the good fit we obtain at $\sim$1600K is coincidental; instead, a different \teff model, that we have ruled out due to overluminosity in the $H$-band, is actually the appropriate one.  We do not think this is plausible, for two reasons.  First, test fits to the secondary with the 1.5--1.7$\mu$m region masked out (not shown) yield the best match to the data at very similar \teff (1650--1700K).  Second, as we will show, the 1600K models also well reproduce all the colors of the secondary from $\sim$$I$ to $L'$ (\S7.4).  Nevertheless, we emphasize that {\it if} our \teff for 2M1207B from $HK$-fitting is skewed by a spurious model overluminosity in the $H$-band peak, then the true \teff of the secondary must be {\it higher} than we estimate, not lower.  As the DUSTY fits in Fig. 10 show, the 1500K DUSTY models, while giving an adequate fit to the data in $K$,  are already {\it fainter} than the data in $H$ rather than brighter.  Going to lower \teff only exacerbates this effect since DUSTY models become redder with decreasing temperature.  At higher \teff, on the other hand, the trend is reversed as the DUSTY models become bluer (e.g., see 1800K model in Fig. 10); normalizing them to match the data in $K$ would indeed make them overluminous in $H$.  If this is due to model opacity problems in $H$, then \teff may be higher than our estimate (though not much higher: for \teff $\gtrsim$ 1900K, not shown, the DUSTY models also deviate from the observed \h2o absorption at $\leq$1.5$\mu$m).  A similar conclusion also holds for the SETTL fits.  While \teff may be somewhat higher our estimate from the $HK$ fits, if model problems exist in the $H$-band, \teff lower than $\sim$1500K is ruled out: regardless of model opacities in the $H$-band, the data do not show the strong $K$-band CH$_4$ absorption that appear in the SETTL models at $<$1500K.  We conclude that, at the least, the $HK$ spectral synthesis sets a strong lower limit of $\sim$1500K on the effective temperature of 2M1207B.  

Finally, we point out that the temperatures we have inferred for 2M1207A and B, 2550$\pm$150K and 1600$\pm$100K respectively, are fully consistent with their approximate spectral types (late-M and mid-to-late L), at least when compared to field dwarfs of similar type with empirically determined \teff (Golimowski et al. 2004).  This is discussed further in \S8.3.  

\subsection{Mass from Age-\teff}
In Fig. 12, we plot the latest Lyon theoretical models (Chabrier et al. 2000) for the temporal evolution of substellar temperatures.  The isochronal age of TWA is $\sim$8$^{+4}_{-3}$ Myr (Song et al. 2003; Zuckerman \& Song 2004; Chauvin et al. 2004).  Combining this with our derived \teff for 2M1207A and B, and comparing to the Lyon evolutionary predictions, yields $M_A$ $\approx$ 18--30 \mj and $M_B$ $\approx$ 6--10 \mj.  We recall from \S5 that the time-evolution of global properties such as \teff in the evolutionary models is insensitive to the particular atmospheres used (COND or DUSTY).   Accordingly, our mass estimates (once \teff is determined) are also nearly independent of the choice of atmosphere in the Lyon evolutionary models.  

On the other hand, there are certainly differences between the evolutionary models constructed by different groups, due to varying assumptions about initial conditions and interior physics.  In particular, for ages of 5--10 Myr, models by the Tucson group (Burrows et al. 1997) predict that masses $\lesssim$ 10 \mj are 5--10\% (100--150K) cooler and 20--30\% fainter than in the Lyon models.  Conversely, masses of $\sim$15--50 \mj are 5--10\% hotter and 20--40\% brighter in the Tucson models than in the Lyon ones (see Burgasser 2004 for more on Lyon/Tucson differences).  Consequently, given our derived \teff and adopted age for 2M1207AB, the Tuscon tracks imply a mass for 2M1207A lower than our Lyon estimate by $\sim$3 \mj, and a mass for 2M1207B higher than our Lyon estimate by $\sim$ 1 \mj: systematic but small shifts compared to the mass uncertainties arising from \teff and age error bars \footnote{The numbers given here for the Tucson tracks are calculated with the web-based calculator, based on Burrows et al. (1997), supplied by the Tucson group at http://zenith.as.arizona.edu/$\sim$burrows/}.    

Our Lyon mass for the primary, $M_A$ $\approx$ 24$\pm$6 \mj, is in good agreement with that inferred earlier by Gizis (2002) and Mamajek (2005).  However, our secondary mass, $M_B$ $\approx$ 8$\pm$2 \mj, is significantly higher than asserted in previous studies (Chauvin et al. 2004, Song et al. 2006: $\sim$5 \mj; Mamajek 2005: 3--4 \mj).  This is the key point of our work, and it is {\it not} affected by the choice of Lyon versus Tucson tracks.  The previous estimates cited here were also based on the Lyon models, like ours; employing Tucson tracks instead shifts all the inferred masses systematically, while leaving unchanged the {\it relative} differences between our mass estimates and previous ones.  We will return, in \S7.4, 7.5 and 8, to the reasons for this discrepancy in the mass of 2M1207B.  For now, we proceed to obtain a second set of mass estimates from color-color comparisons.   

\subsection{Mass from Colors}
Fig. 13 shows Lyon theoretical isochrones for substellar masses, at two ages (5 and 10 Myr) bracketing the expected range for TWA, and in two color-color planes: $J$-$H$ versus $H$-$K$ and $H$-$K$ versus $K$-$L'$.  Isochrones are shown for both COND and DUSTY cases (as mentioned in \S5, the particular atmosphere adopted as outer boundary condition in the evolutionary models does matter when examining the emergent SED).  Notice that for masses $\gtrsim$ 30 \mj, the COND and DUSTY isochrones become identical, because at the \teff corresponding to these masses at 5--10 Myrs, no photospheric dust has formed yet.  At smaller masses (i.e., lower \teff), burgeoning grain formation causes the COND and DUSTY isochrones to diverge, with DUSTY models becoming substantially redder in all the infrared colors.  Comparing these to the observed colors of 2M1207A and B, we find the following. 

{\it Primary}: 2M1207A lands perfectly on the 5--10 Myr isochrones in the [$H$-$K$]--[$K$-$L'$] plane.  The implied mass is 20--30 \mj, in complete agreement with the value derived above from age-\teff considerations.  While COND and DUSTY are almost indistinguishable at these masses, close examination reveals a slightly better match to DUSTY, as expected when grains first start to appear.  On the other hand, the primary is somewhat offset from the isochrones in the [$J$-$H$]--[$H$-$K$] plane.  The absence of such an effect in $[H-K]$--$[K-L']$ suggests a discrepancy with the model $J$-$H$ alone.  Indeed, shifting the Lyon isochrones to redder $J$-$H$ by 0.2 mag produces a good match to the primary, again at 20--30 \mj.  A $\sim$0.2 mag offset is also seen in $J$-$K$, in a [$J$-$K$]--[$H$-$K$] plot (not shown).  In fact, an analogous $J$-$K$ offset is known to exist between the Lyon models and field M dwarfs, with the former again being too blue by 0.2 mag (Allard et al. 2000, 2001).  As the latter authors discuss, this may be due to remaining incompleteness, at relatively warm temperatures, in the AMES-\h2o linelists used in the synthetic spectra (FeH opacities may also play a role; see \S7.4).  Given the qualitative and quantitative correspondence between our and the field M dwarf results, we are justified in applying a +0.2 mag correction to the Lyon isochrones in $J$-$H$ (or equivalently, in $J$-$K$) for this source.  The final outcome is that the same mass of 20--30 \mj is implied for 2M1207A from both color-color diagrams and the age-\teff comparisons.  

{\it Secondary}: The observed colors of 2M1207B are in very good agreement, within the errors, with the 5--10 Myr DUSTY isochrones in both [$J$-$H$]--[$H$-$K$] and [$H$-$K$]--[$K$-$L'$] planes.  The implied mass is $\sim$6--11 \mj, identical to that derived from the age-\teff model comparisons.  Notice that the secondary lies very far from the COND isochrones in these color-color plots, bolstering our earlier conclusion that grains remain suspended in its atmosphere.
  
\subsection{Effective Temperature from Colors}

The agreement in mass between the color-color and age-\teff techniques fundamentally means that: the synthetic spectra that best fit the observed $HK$ spectra at a particular \teff also faithfully reproduce the shape of the SED (i.e., colors) all the way from $J$ to $L'$.  

This is explicitly illustrated in Figs. 14a,b.  We earlier derived spectroscopic \teff = 2550$\pm$150K and 1600$\pm$100K for 2M1207A and B.  The plots show the DUSTY colors for these \teff (at log $g$ $\approx$4.0, as deemed appropriate by the Lyon tracks), compared to the data for the primary and secondary.  Not surprisingly (since the \teff were determined from spectral fits in the $HK$ region) the synthetic colors match the observed $H$-$K$ for both objects.  In addition, as expected, the models fit reasonably well the $J$-$H$ and $K$-$L'$ colors in both, especially after correcting for the known systematic offset of 0.2 mag in the model $J$-$H$ described earlier.  We note that for 2M1207B, the $JHKL'$ colors are best matched by 1550$\pm$50K: slightly lower in the mean, but within the 1600$\pm$100K range from our $HK$ spectral fits.   

For both objects, we also plot in Figs. 14a,b the HST [F090M-F160W] colors observed by Song et al. (2006), corresponding roughly to $I$-$H$.  For 2M1207A, we see that the DUSTY model at the spectroscopic \teff $\sim$ 2550K appears too blue, by $\sim$0.3--0.4 mag, compared to the HST data.  The [F090M-F160W] color is best matched instead at 2250K, cooler than the 2550K$\pm$150K models that fit both the $HK$ spectra and the $JHKL'$ colors.  An analogous offset of $\sim$0.3 mag, in $I$-$K$, is also seen for field mid-M dwarfs in the data presented by Leggett et al. (2000; see their Fig. 12).  Using the latter sample, Mohanty et al. (2004b) further show that the offset appears to arise from a model overluminosity in $I$ compared to the observations.  Remaining opacity uncertainties in the synthetic spectra probably account for this.  In particular, the F090M filter covers 0.8--1.0$\mu$m; in mid-to-late M and early L dwarfs, this entire region exhibits strong FeH opacity due to the ${F^4}{\Delta}_i$-${X^4}{\Delta}_i$ transition (the most prominent band of which is the 0--0 Wing-Ford bandhead at 0.9896$\mu$m; Wing \& Ford 1969; Phillips et al. 1987; Kirkpatrick et al. 1999; Dulick et al. 2003).  The DUSTY spectra (or any Lyon synthetic spectra for that matter), however, do not yet include any FeH opacity at $\lesssim$1 $\mu$m (F. Allard, 2006, pvt. comm.), since oscillator strengths for this transition were unknown till recently. Dulick et al. (2003) have now calculated detailed line strengths and opacities for this transition, from high level ab initio calculations of the electronic dipole transition moment.  Inclusion of these new opacities could conceivably resolve the F090M (or equivalently, $I$-band) discrepancy observed in 2M1207A and field M dwarfs relative to the synthetic spectra\footnote{An overluminosity in $I$ is also seen in the COND models compared to T dwarfs; this is ascribed to uncertainties in the model treatment of H$_2$ and He collisional broadening of the far line wings of the \na and \pot alkali doublets (Allard et al. 2001; Baraffe et al. 2003).  However, this appears unlikely to cause the $I$-band discrepancy in 2M1207A and field M dwarfs: the alkali doublets are much narrower in these sources than in the high-pressure atmospheres of T dwarfs.  FeH uncertainties seem a better bet in this case.}.  Moreover, FeH lines due to the ${F^4}{\Delta}_i$-${X^4}{\Delta}_i$ transition extend into the $J$-band as well (most prominently, bandheads of the 0--1 and 1--2 bands at 1.1939 and 1.2389$\mu$m, respectively).  It is thus possible that the $J$-band offset noted earlier in 2M1207A and field M dwarfs could also be fixed with the new opacities (since the FeH $J$-band opacities currently in the Lyon spectra are only very rough initial estimates; F. Allard, 2006, pvt. comm.).  At any rate, given the qualitative and quantitative similarity between our results for 2M1207A and the field M dwarfs, we are justified in postulating a $\sim$ +0.3 mag redward offset in the DUSTY [F090M-F160W] colors; this makes the 2550$\pm$150K models agree as well with the 2M1207A HST data as they do with its $JHKL'$ colors and $HK$ spectrum.

For 2M1207B, we see that the observed [F090M-F160W] color is in very good agreement with the 1550--1600K DUSTY models, consistent with its other colors and our spectroscopic \teff of 1600$\pm$100K.  The absence of any obvious offset in 2M1207B's [F090M-F160W] color from the model predictions, unlike the situation in 2M1207A, may indicate that FeH is indeed the culprit in the primary as we suggest above, since the effect of FeH opacities should weaken by the mid-to-late L type of the secondary.   

In summary, our analysis so far yields a {\it self-consistent} picture of colors, \teff, age and mass for 2M1207A and B: model spectra and Lyon evolutionary tracks for age = 5--10 Myrs, \teff$_A$ = 2550$\pm$150K, \teff$_B$ = 1600$\pm$100K, $M_A$ $\approx$ 24$\pm$6 \mj and $M_B$ $\approx$ 8$\pm$2 \mj agree with all the spectroscopic and color data available, especially after accounting for known remaining uncertainties in model opacities.  The only other question is: do the evolutionary models reproduce the absolute photometry and luminosities of both components as well?

\subsection{Absolute Magnitudes, and Mass from Age-Luminosity}

Using Mamajek's (2005) distance estimate of 53$\pm$6 pc, the absolute photometry of 2M1207A and B can be computed from their apparent magnitudes.  Mamajek then combined the resulting M$_K$ with the empirical $K$-band bolometric corrections ($BC_K$) for field dwarfs (Golimowski et al. 2004), to derive the bolometric luminosities of 2M1207A and B (the errors associated with using field dwarf $BC_K$ for these young objects are adressed further below).  We compare the resulting M$_{F090M,J,H,K,L'}$ and $\mbol$ to the Lyon evolutionary predictions with DUSTY atmospheres \footnote{Chauvin et al. (2004) followed the same procedure as Mamajek (2005), but with a larger distance of 70pc; since both Mamajek and Song et al. (2006) find a smaller distance with improved data, we do not consider the Chauvin values here. Moreover, since Song et al.'s distance of 59$\pm$6 pc is nearly identical to Mamajek's, our results are the same for either value; we consider only the Mamajek distance here for clarity.  We will discuss the Chauvin et al. and Song et al. results at the appropriate points.}.  Note that we include only the F090M ($\sim$ $I$-band) photometry from Song et al.'s (2006) HST observations, since their F110M and F160W points are roughly equivalent to the $J$ and $H$ photometry already in our dataset.  The results are plotted in Figs. 14d,e, and reveal the following.

{\it Primary}: Fig. 14d shows the emergent SED according to the Lyon DUSTY evolutionary models, for an object at our preferred \teff = 2550K and age = 5 and 10 Myr, compared to M$_{F090M,J,H,K,L'}$ and M$_{bol}$ for 2M1207A with $d$ = 53 pc.  We see that the 10 Myr Lyon model matches the absolute $HKL'$ photometry and bolometric luminosity of 2M1207A remarkably well: to better than 0.05 mag in $HKL'$, and better than 0.15 mag in M$_{bol}$.  A slightly larger deviation of $\sim$0.2 mag is seen in M$_J$; similarly, a 0.4 mag deviation appears in M$_{F090M}$.  These are precisely the offsets that cause the models to be systematically bluer than the data in $J$-$H$ and F090M-F160W, in the color comparisons in Fig. 14a.  As discussed in \S7.3 and 7.4, these arise from known incompleteness in the \h2o and FeH opacities and are seen in field M dwarfs as well.  As such, these deviations are not a significant cause for concern in our present analysis.  For example, Fig. 14d shows that a \teff = 2450K model matches the observed M$_{F090M}$ and M$_{J}$ very well, while being slightly fainter than the data (by $\sim$0.2 mag) in M$_{H,K,L'}$ and M$_{bol}$: a small shift of only 100K from our preferred \teff.  For the same distance, the 5 Myr model is brighter in all bands and overall luminosity by $\sim$0.2 mag; however, this age can also be accomodated by a $\sim$5 pc increase in distance, within the errors of the Mamajek's determination. On the other hand, the $d$ = 70 pc used by Chauvin et al. (2004) makes the primary brighter by 0.6 mag and significantly degrades the model fit at either age (not shown).  Since Mamajek's $d$ = 53$\pm$6 pc is based on space motions alone, the excellent fit it yields to the Lyon evolutionary models in our {\it independent} luminosity analysis  makes it (and equivalently, Song et al.'s very similar estimate) the preferred distance.  Changing \teff by the $\pm$150K uncertainty in our derivation does not appreciably alter these conclusions.  Note that the associated Lyon model radius for 2M1207A, for \teff $\approx$ 2550K and age = 5-10 Myr, is $\sim$0.26 \rsun.  This agrees very well with the 0.24 \rsun inferred directly from its \lbol and \teff.  

{\it Secondary}:  Fig. 14e depicts the same analysis for 2M1207B, using Lyon DUSTY evolutionary models at 5 and 10 Myr with our preferred \teff = 1600K.  As expected from our prior color analysis, the Lyon models at either age reproduce the observed {\it shape} of the secondary's SED quite well (small deviations from the shape, apparent in F090M and $L'$, are discussed in \S9).  However, the secondary appears {\it considerably fainter than predicted, by roughly the same amount in all photometric bands and bolometric luminosity}.  For our preferred $d$ = 53 pc, it is fainter than the models by 2.5$\pm$0.5 mag in $F090M,J,H,K,L'$ and $\mbol$.  Even for Chauvin et al's overestimated $d$ = 70 pc, the deviation would still be 1.9$\pm$0.5 mag.  Varying \teff by the $\pm$100K uncertainty in our derivation has a negligible impact on this conclusion.  

On the other hand, for $d$ = 53 pc and age 5--10 Myr, \teff $\approx$ 1000K Lyon models provide a much better match to the secondary's bolometric luminosity, corresponding to masses of 3--4 \mj.  This is illustrated by the 1000K (4 \mj at 10 Myr) model plotted in Fig. 14e.  For Chauvin et al.'s $d$ = 70 pc, a similarly good match to $\mbol$ is found with \teff $\approx$ 1250K (5 \mj at 10 Myr; not shown).   This is not surprising, since the 3--5 \mj estimates by Chauvin et al. (2004) and Mamajek (2005) were {\it based} on age-$\mbol$ analyses.  However, models at these low \teff, unlike those at $\sim$1600K, fail miserably at reproducing the observed {\it shape} of the secondary's SED, as discussed in our earlier \teff and color-color analyses.  For example, note the predicted SED for the 1000K DUSTY Lyon model in Fig. 14e: far redder, and 3--5 mags fainter in $J$ and $H$ and 7 mags fainter in F090M, than observed (equally large deviations occur for COND or SETTL models at similar \teff, with the models now being too blue and much brighter than observed in $J$ and $H$; not shown in Fig. 14e, but see discussion in \S 7.1).  Conversely, matching the observed absolute magnitude in, say, F090M, requires a \teff = 1450K model (see Fig. 14e), which again fails entirely in fitting the SED at longer wavelengths, and is also brighter than the observed $\mbol$ by 2 mags.    

In summary, there is a stark contrast between 2M1207A and B.  For the estimated age and distance of the system, the \teff indicated by the primary's spectrum and colors also adequately reproduces its absolute photometry and bolometric luminosity.  In the secondary, however, there is a serious disagreement between the \teff suggested by its observed luminosity and the \teff implied by its spectrum and colors.  

\subsection{Comparison to AB Pic B}

To drive home the point that 2M1207B is indeed deviant, we have also performed the same analysis for the recently discovered young, very low mass brown dwarf AB Pic B (Chauvin et al. 2005b).  The spectral type and age of this object ($\sim$L1--L3, 30 Myr), as determined by the latter authors, are roughly similar to that of 2M1207B; as such, it serves as an excellent test of our 2M1207B analysis.  Our results are plotted in Figs. 14c,f.

Fig. 14c shows the $J$-$H$ and $H$-$K$ colors reported for AB Pic B by Chauvin et al. (2005b; we have converted to CIT from their reported 2MASS filters, as discussed in \S5), compared to DUSTY and SETTL models at log $g$ = 4.25 (the gravity indicated by the Lyon evolutionary models for brown dwarfs at $\sim$30 Myr).  We see that the observed $J$-$H$ and $H$-$K$ are consistent with DUSTY models at 1600--1700K and 1800K respectively, overall suggesting \teff $\sim$ 1700K.  Indeed, the 1700K SETTL model, which is slightly bluer (by 0.2 mag) in $H$-$K$ than the DUSTY model at the same \teff, clearly provides a good match to both the observed $J$-$H$ and $H$-$K$.  We therefore adopt \teff $\approx$ 1700$\pm$100K; a temperature that is very similar to the 1600$\pm$100K derived for 2M1207B.  

The bluer $H$-$K$ in SETTL, compared to DUSTY, in the above fits deserves a word.  At log $g$ = 4.0, SETTL and DUSTY spectra predict very similar $H$-$K$ for \teff $\gtrsim$ 1600K; this is why we obtained nearly identical results from our earlier DUSTY and SETTL fits to the $HK$ spectrum of 2M1207B.  In the AB Pic B color fits, however, we employ a slightly higher gravity of log $g$ = 4.25.  As discussed in \S5, the mixing scale-height in the SETTL models is proportional to $\sqrt{1/g}$, and thus decreases (implying more settling) with increasing gravity.  Consequently, for a given \teff, the SETTL models move closer to COND ones as gravity goes up, and thus become increasingly blue in $H$-$K$ compared to DUSTY (for the reasons described in \S7.1.1).  This is in line with observations, as follows.  As noted earlier, high-gravity field L dwarfs with \teff similar to 2M1207B exhibit considerable grain settling, while the much younger and lower gravity 2M1207B evinces hardly any settling at all, thus appearing redder in $H$-$K$ than the latest field L types.  AB Pic B, again at a similar \teff but slightly older and higher gravity than 2M1207B, fits nicely into this sequence by showing a hint of settling and thus somewhat bluer $H$-$K$ than 2M1207B.  
            
We now compare the absolute $JHK$ photometry and bolometric luminosity of AB Pic B, computed for the estimated $d$ = 45.5 pc to the AB Pic system (Song et al. 2003, from $Hipparcos$ \footnote{Chauvin et al. (2005b) quote a slightly larger distance of 47.3 pc, but do not give a reference for this value.  We thus adopt the $Hipparcos$ distance cited by Song et al. (2003).  The small difference between the two values has no effect on our results.}), to the Lyon DUSTY evolutionary predictions for \teff = 1700K, age = 30 Myr.  Ideally, given our color results above, we should compare to Lyon models with SETTL instead of DUSTY atmospheres, but these have not yet been constructed.  However, the precise atmosphere used makes very little difference to the predicted bolometric luminosity, as stated before.  Moreover, the difference in predicted absolute photometry between the SETTL and DUSTY models will only be at the $\sim$0.2 mag level indicated by the $H$-$K$ color difference in Fig. 14e.  For our purposes here, therefore, the available Lyon DUSTY models will suffice.  The results are plotted in Fig. 14f.  We see that the model predictions match the data extremely well, to within 0.1 mag in M$_J$, M$_H$ and $\mbol$ and 0.2 mag in M$_K$.  The slightly redder model slope from $H$ to $K$, compared to the data, is precisely the difference that should be resolved by using a SETTL instead of DUSTY atmosphere.  

We point out that Chauvin et al. (2005b) have also inferred \teff $\sim$ 1700K for AB Pic B, by directly comparing its observed bolometric luminosity to that predicted by the Lyon evolutionary models for various \teff at the system's age and distance.  Our analysis has proceeded from a more fundamental level.  We have first derived \teff $\sim$ 1700K by comparing the observed colors to synthetic spectra, and then shown that the Lyon evolutionary model for this temperature also matches the observed absolute photometry and bolometric luminosity.  Our results for AB Pic B mirror those for 2M1207A; neither of these sources evinces any sign of the large deviation between models and data seen in 2M1207B.

These results are encapsulated in Fig. 15, where the data are placed on the Lyon theoretical H-R diagram using $d$ = 53 pc and 45.5 pc for 2M1207AB and AB Pic B respectively. For 2M1207A, the age-\teff, age-\lbol and \tefflbol comparisons to the models yield completely self-consistent results: \teff $\approx$ 2550$\pm$150K, log [L/\lsun] = 2.68$\pm$0.12, mass = 20--30 \mj and age = 5--10 Myr.  Similarly, for its observed bolometric luminosity and our derived \teff $\approx$ 1700, AB Pic B lands nicely on the $\sim$30 Myr isochrone, in complete agreement with its expected age (the associated mass is $\sim$13 \mj).  For 2M1207B, however, the results are highly divergent.  Fixing age-\teff yields mass 6--10 \mj, but predicts \lbol 8 times higher than observed; fixing age-\lbol gives mass 3--4 \mj but \teff 600K lower than we derive, and fixing \tefflbol predicts age $>$0.5 Gyr, much older than expected for a TWA member.  The agreement between models and data for 2M1207A, and particularly for AB Pic B, which is roughly similar in age, spectral type and temperature to 2M1207B, strongly suggests that our analysis is accurate, and that 2M1207B is uniquely deviant for a real physical reason.    

Finally, we note that our adopted \lbol for all objects are derived using $BC_K$ appropriate for field dwarfs.  In particular, we use mid-to-late L field dwarf $BC_K$ for 2M1207B.  The spectrum and colors of 2M1207B, however, are much redder than in these dwarfs; i.e., its $K$-band flux represents a larger fraction of its total bolometric flux than in the field objects.  This effect can be seen clearly in the COND/DUSTY model comparisons at various gravities shown previously in Fig. 6.  All the models here have been normalized to unit area by dividing by the appropriate bolometric flux $\sigma$T$_{eff}^4$.  We see that for a given \teff $\lesssim$ 2000K, the relative $K$-band flux increases with decreasing gravity, in both COND and DUSTY.  Moreover, for a given gravity and \teff, the $K$-band flux in the DUSTY models is larger than in the COND ones.  These effects arise from the various opacity issues discussed in \S7.1.1.  Now, we have found that the colors and spectrum of 2M1207B are well matched by a low-gravity DUSTY model.  Conversely, field mid-to-late L dwarfs at \teff similar to 2M1207B have high gravities, as well as enhanced settling corresponding more closely to COND models.  Consequently, the $K$-band flux in 2M1207B, relative to its bolometric flux, should indeed be higher than in field L dwarfs, due to both gravity and settling effects.  Applying the dwarf $BC_K$, therefore, overestimates 2M1207B's observed luminosity; correcting for this should make it appear {\it even fainter} than the theoretical prediction, exacerbating all the discrepancies cited above. 

\section{Discussion}
We consider five possible explanations for the anomalous behaviour of 2M1207B: 

{\noindent}Our \teff is correct, but the theoretically predicted luminosity is too high because \\{(\it 1)} 2M1207B is much further away or substantially older than we assume, or \\{(\it 2)} its radius is smaller than the evolutionary models predict for its temperature and age (i.e., the theoretical \tefflbol relationship is wrong).

{\noindent}{\it Or} the age, distance and theoretical \tefflbol relationship are correct, but \\{(\it 3)} the true \teff is lower than the spectral models suggest, due to opacity errors in the latter.

{\noindent}{\it Or} the age and distance are correct, but our \teff and \lbol are inappropriate because \\{(\it 4)} this is an intrinsically bluer, brighter source, heavily {\it reddened} by dust in our line of sight.

{\noindent}{\it Or} the age, distance and inferred \teff are all accurate, but 2M1207B appears underluminous \\{\it (5)} due to {\it gray} extinction by a nearly edge-on disk.

{\noindent}We consider each possibility in turn, and show that only the last is viable.  
  
\subsection{Distance and Age Uncertainties}

The young age (5--10 Myr) of TWA is supported by various independent lines of evidence: presence of Lithium absorption, large X-ray emission and strong chromospheric activity in its members, location of its members in the H-R diagram, and its inclusion of at least some actively accreting classical T Tauri stars (Kastner et al. 1997; Webb et al. 1999; Muzerolle et al. 2000; Zuckerman et al. 2001; Weinberger et al. 2002; Song et al. 2003; Mohanty et al. 2003; Zuckerman \& Song 2004).  The nearby distance ($\sim$40--70 pc) to the association is supported by both trigonometric parallaxes to a few members (Perryman et al. 1997; Wichmann et al. 1998) and a moving cluster analysis based on individual space motions (Mamajek 2005).  As discussed in \S1, 2M1207A exhibits several characteristics of youth consistent with TWA membership.  The data reported in this paper -- both the peaked $H$-band spectral shape and weak $K$-band \na absorption indicative of low gravity, and the excellent agreement with the theoretical evolutionary tracks for an age of 5--10 Myr and $d$ $\approx$ 53 pc -- provide further evidence of its youth and proximity.  

{\it If} 2M1207B is physically associated with 2M1207A, therefore, it cannot be further away than $\sim$50--70 pc or significantly older than 5--10 Myr.  Explaining its nearly order of magnitude luminosity discrepancy through distance/age variations, however, requires grossly larger values: $d$ $\sim$ 150--200 pc and/or age $>$ 0.5 Gyr.  Since the astrometry presented by Chauvin et al. (2005a) and Song et al. (2006) imply, to a very high significance, that 2M1207A and B indeed form a single bound system, such a large distance or age can be ruled out.  

\subsection{Evolutionary Model Uncertainties}

Alternatively, for our adopted age and \teff (and hence model-implied mass) for 2M1207B, might the radius (and thus luminosity) predicted by the evolutionary tracks be too high?  Its true radius must then be $\sim$3 times smaller than expected, to produce the requisite reduction in luminosity.  We examine a suite of recent results to test the viability of this hypothesis.  

From high-resolution spectroscopic analysis of \teff and gravities, Mohanty et al. (2004a, b) conclude that at very young ages (few Myr), the theoretical radii are in good agreement with observations down to masses of about 0.03 \msun, but possibly too small by up to a factor of $\sim$2 in the lowest mass brown dwarfs.  These results remain to be verified through more accurate (and empirical) radius measurements; more importantly, the implied trend -- true radii {\it larger} than predicted -- is in the opposite sense to what we require.  

On the other hand, Close et al. (2005) have suggested that AB Dor C, with a dynamical mass of 0.09 \msun, is $\sim$2.5 times fainter than predicted, for their adopted age of 50 Myr.  Prima facie, this discrepancy is in the direction we require.  However, while Close et al.'s inferred mass is robust, their claim of a luminosity offset from the tracks is highly sensitive to their assumed age.  Detailed analyses by Luhman et al. (2005) show that 50 Myr is too low: the true age appears to be $\sim$100 Myr, removing the postulated luminosity offset.  

Finally, all current empirical tests of the mass-radius relationship in stars, brown dwarfs and planets -- interferometric radius measurements of $\sim$0.1--0.5 \msun field M dwarfs (S\'{e}gransan et al. 2003); eclipsing binary measurements for higher mass stars ($\sim$0.5--0.8 \msun; Torres \& Ribas 2002); very recent measurements for the first young eclipsing binary brown dwarf system (in Orion, age $\sim$ 1 Myr, $m$ = 0.06 and 0.03 \msun; Stassun et al. 2006); and radius measurements for transiting hot Jupiters (Baraffe et al. 2005) -- indicate that the theoretical radii are correct to within 25\%.  Moreover, in those cases where offsets exist between the models and data, the true radii are again {\it larger} than predicted (for various physical reasons in different mass/age/environment regimes; Baraffe et al. 2002, 2003; Chabrier et al. 2004).  

In summary, all reliable evidence, from stellar to planetary masses over a range of ages, points to generally reasonable agreement between the observed and theoretical radii for a given mass; when the models err, it is by predicting sizes that are somewhat too small, instead of too big.  In this light, a very large (factor of 3) {\it over}estimation of radius by the evolutionary tracks for 2M1207B appears untenable.  

The above arguments apply if 2M1207B formed in a manner similar to an ``isolated body''; i.e., conventional brown dwarf evolutionary models are appropriate.  However, Fortney et al. (2005) suggest that planets formed via core-accretion should be considerably smaller and fainter in their youth than standard evolutionary models predict, due to initial condition effects.  While this effect may not be apparent in the transiting exoplanets discovered so far and discussed above, since they all have ages $\gtrsim$ few 100 Myrs, can it explain our results for the much younger 2M1207B?  The answer is no: as Fortney et al. themselves point out, 2M1207B is not expected to have formed by core-accretion.  Calculations by Lodato et al. (2005) show why.  For {\it in situ} formation of 2M1207B at its projected separation from the primary ($\sim$40 AU for $d$ $\sim$ 53 pc), in the available time ($\lesssim$ 10 Myr), the required disk surface density for core-accretion translates to a disk mass far too large to be entertained.  Core-accretion much closer to the primary followed by outward migration is also not possible, since plausible timescales and disk masses require formation at disk radii $<$0.6 AU, which would cause {\it inward} migration.  With formation by core-accretion ruled out, we find no compelling physical rationale for postulating a grossly overestimated radius for 2M1207B.

Lastly, is it possible that the predicted luminosity for 2M1207B is too high due to errors in the theoretical mass-\teff relationship, instead of in mass-radius?  In particular, consider the implications if the predicted mass-radius relation is fairly accurate (as the studies above strongly suggest), but the mass-\teff relation is too cool; i.e., for a given mass and age, the evolutionary tracks indicate the correct radius but too low a temperature.  For an {\it empirically} determined mass, of course, this would underestimate the true luminosity.  However, we {\it derive} the mass of 2M1207B from age-\teff comparisons to the theoretical tracks, using a temperature independently inferred from fits to synthetic spectra.  Assuming our \teff is in the right ballpark (as we will argue in \S8.3), and our adopted age is valid, a mass-\teff relationship that is too cool will lead us to derive too high a mass to match our (correct) \teff, and thus too large a radius.  Consequently, the predicted luminosity will be too high; can this explain our results? The answer is no.  Our mass for 2M1207B is already below the Deuterium-burning boundary ($\sim$12 \mj), and will be even lower if, as implied in the above scenario, the mass is actually overestimated.  At an age of 5--10 Myr, the theoretical mass-radius relationship indicates that objects near and below the $D$-burning boundary all have very similar radii, with masses from 0.5 to 15 \mj differing by $<$50\% in radius (Baraffe et al. 2003).  Hence the luminosity will be overestimated by no more than a factor of $\sim$2, far too small to explain the nearly order of magnitude discrepancy seen in 2M1207B.  This conclusion becomes even stronger if, as Mohanty et al. (2004b) and Stassun et al. (2006) suggest, the theoretical radii at young ages are somewhat underestimated for low substellar masses, since this would make the luminosity overprediction in the above scenario even smaller.

In conclusion, radius or \teff errors in the evolutionary models are not a plausible explanation for 2M1207B appearing vastly fainter than predicted.  From an empirical standpoint, the remarkably good match between the theoretical and observed luminosities in both 2M1207A and AB Pic B also argues strongly against serious problems in the evolutionary models.  The evidence from AB Pic B is especially compelling, since it has a roughly similar spectral type and nearly the same \teff, from synthetic color comparisons, as 2M1207B, while being only slightly older.  Consequently, its expected mass and luminosity from age-\teff model comparisons are close to the values predicted for 2M1207B.  As such, it is hard to imagine a missing piece of evolutionary physics that generates a very large difference between the predicted and observed luminosities in 2M1207B, but causes no such offset in AB Pic B.  

\subsection{Atmospheric Model Uncertainties}
Another possibility is that the theoretical \tefflbol relationship is correct, but it is the atmospheric models that are wrong: erroneous opacities cause us to infer, from both our spectral and color analyses, \teff too high by $\sim$600K.  Without other young ultra-cool objects with {\it empirically} determined \teff, we cannot test this proposition directly.  However, four lines of evidence argue strongly against such large errors in the atmospheric models.

The first is a qualitative argument.  Matching the observed bolometric luminosity of 2M1207B, at its age and distance, requires \teff $\sim$ 1000K (\S7.5).  Conversely, atmospheric models over a narrow but much higher range of \teff, 1600$\pm$100K, reproduce very well both its $HK$ spectral shape and, independently, its $\sim$$I$ to $L'$ colors.  This large wavelength coverage includes opacities from a number of different sources, e.g., \h2o, H$_2$-CIA, TiO and dust grains.  Errors in all these opacities would have to be very finely tuned indeed, for the models at $\sim$1600K to fortuitously reproduce precisely the same spectral shape and colors appropriate to a $\sim$600K lower temperature, over the entire $IJHKL'$ range.  Instead, any severe opacity errors are more likely to yield very {\it divergent} \teff estimates from the different color and spectral comparisons.  This is not seen in our analysis, and suggests that the synthetic spectra are not too far off the mark in their temperature prediction.
  
An examination of $K$-$L'$ colors provides a second, more compelling argument.  The $K$-$L'$ colors of field  M, L and T dwarfs are known to form a tight monotonically increasing sequence with decreasing \teff, despite the range in age and dust properties encompassed by these sources (Golimowski et al. 2004).  This is illustrated in Fig. 16, where we plot $K$-$L'$ in MLT dwarfs against their semi-empirically derived `structural' \teff (data from Golimowski et al. 2004; \teff via parallax, flux-calibrated spectra and theoretical \teff-\lbol relationship; see latter paper for details)\footnote{From the Golimowski et al. sample, we have excluded those sources $(a)$ known (or suspected by Golimowski et al.) to be unresolved binaries, since the \lbol, \teff and colors of the individual components are not well-determined, and those of the combined systems are prone to systematic offsets from single dwarfs (Liu \& Leggett 2005); $(b)$ for which Golimowski et al. estimate $L'$ flux by extrapolation from $JHK$ photometry and spectral type, since these estimates have large errors, leading to higher uncertainties in \lbol and \teff than in sources with directly measured $L'$ flux; and $(c)$ for which NIR and optical spectral types differ by $>$ 1 sub-type, when classifications are available in both, since these may be anomalous.}.  The \teff errors are $\sim$ $\pm$150K for sources with imprecise ages, and $\pm$100K for better constrained ages\footnote{For dwarfs with imprecise ages, Golimowski et al. conservatively assume 0.1--10 Gyr.  For a handful of the dwarfs in resolved multiple systems, ages are better constrained by assuming coevality with the Main Sequence primaries.  These are GD 165B: 1.2--5.5 Gyr, Gl 584C: 1--2.5 Gyr, Gl 570D: 2--5 Gyr, and Gl 229B: 0.5--10 Gyr (Golimowski et al. 2004 and references therein)}.  The plot clearly shows the monotonic color sequence with \teff\footnote{Golimowski et al. (2004) and Leggett et al. (2002a) have noted that $K$-$L'$ levels out over the spectral type range L6--T5, though it rises with later type on both sides of this interval.  However, Golimowski et al. also find that the structural \teff corresponding to L6--T5 is roughly constant ($\sim$1300--1500K).  Consequently, $K$-$L'$ rises much more monotonically with \teff than with spectral type, as evinced by our plot (all the L6--T5 sources are bunched together at nearly the same \teff and $K$-$L'$; green points and asterisk in our plot).  As Golimowski et al. discuss, the constancy of both \teff and $K$-$L'$ over the L6--T5 interval are probably related, arising from dust-settling and cloud-clearing phenomena in these field L-T transition objects.}.  We emphasize that this trend is {\it independent} of any atmospheric modeling.   

Now we examine 2M1207A and B on the same plot.  Their observed $K$-$L'$ colors and the \teff we {\it derive} from the atmospheric models clearly place both sources firmly within the empirical field dwarf sequence.  In particular, adopting \teff lower by $\sim$600K than our estimate would make 2M1207B a significant outlier (even after accounting for the $\pm$150K uncertainty in the field dwarf temperatures).  This further supports our inferred \teff.  Moreover, our \teff for 2M1207A and B -- 2550$\pm$150K and 1600$\pm$100K -- place them precisely in the regions of the diagram expected for their approximate spectral types: late M and mid-to-late L respectively.  This is another indication that our temperatures are in the right ballpark.  

AB Pic B provides the third argument.  Using model atmospheres with log $g$ = 3.5--4.5, we find \teff = 1600$\pm$100K for 2M1207B.  With a gravity in the same range, log $g$ = 4.25, we derive a very similar temperature, $\sim$ 1700K, for AB Pic B as well.  If severe synthetic opacity errors cause us to overestimate \teff in 2M1207B by $\sim$600K, then the same should hold for AB Pic B too; i.e., the latter should also appear severely underluminous compared to the evolutionary tracks.  The fact that it does not bolsters our confidence in the \teff inferred for 2M1207B, since we use the same atmospheric models for the temperature analysis of both.  
 
Our final argument is based on the temperatures of field MLT dwarfs.  A number of of these have \teff estimated from spectral modeling, as well as semi-empirical `structural' \teff derived from their empirical \lbol combined with the theoretical \teff-\lbol relationship for field ages (Dahn et al. 2002; Golimowski et al. 2004; Vrba et al. 2004).  These structural \teff, while still dependent on the evolutionary models, are our best current estimates of the true temperatures of these dwarfs.  Comparing them to the spectroscopic \teff indicates the following systematics in the latter (see also Smith et al. 2003).  For mid-M dwarfs, where dust formation is marginal, the spectroscopic \teff seem systematically too cool by $\sim$200K.  This is ascribed to remaining lacks in the NIR model \h2o opacities at these relatively hot temperatures (Leggett et al. 2001).  Conversely, in the late L to early T dwarfs, where ongoing dust settling is a major phenomenon, the spectroscopic \teff appear too hot by $\sim$200K (e.g., Schweitzer et al. 2002).  This is because nearly all comparisons to these sources have been on the basis of inappropriate no-settling (i.e., DUSTY) or fully settled (i.e., COND) model conditions; new models that do incorporate partial settling (e.g., SETTL) now show marked improvements in this regime (e.g., Allard et al. 2003; Knapp et al. 2004; Burrows et al. 2006).  On the other hand, in late-M to mid-L dwarfs where dust remains fully suspended in the atmosphere, and mid-to-late T dwarfs where dust has entirely settled below the photosphere, the spectroscopic \teff agree quite well with the structural ones, with a scatter of order the $\sim$100--200K uncertainty in the latter (e.g., Leggett et al. 2001; Schweitzer et al. 2001; Allard,N. et al. 2003; Leggett et al. 2002b; Burrows et al. 2006).  Hence, given the very dusty photosphere indicated by the extreme red colors and spectrum of 2M1207B, and the good agreement between spectrospcopic and structural \teff in the fully DUSTY regime, we do not expect a sytematic error in excess of 100--200K in our \teff for 2M1207B.  Morever, we have seen that in both 2M1207A and AB Pic B, the luminosities implied by combining the theoretical \teff-\lbol relationship with our spectroscopic \teff match the observed luminosities very well, within the 100--150K uncertainty in our temperature determination.  This too implies that our \teff systematics are $\lesssim$150K.

The combined weight of the above suggests that, while we cannot rule out 100--200K systematics in our \teff for 2M1207B, due to model opacity uncertainties, it is highly implausible that we are overestimating \teff by the $\sim$600K required to match its observed luminosity.

\subsection{Interstellar Extinction}
Yet another possibility is that 2M1207B is intrinsically bluer and brighter than observed, but heavily reddened by dust in our line of sight.  While this would not affect age or distance, it would nullify the \teff and \lbol we derive assuming negligible extinction and reddening.  

Large interstellar extinction towards a TWA member is extremely unlikely, however.  The association is far from any known molecular cloud that could cause significant reddening; it is also very nearby, implying very little intervening ISM dust.  In keeping with this, the extinction towards the eponymous T Tauri star TW Hydra, as measured from HI absorption in Ly$\alpha$ emission assuming an interstellar gas-to-dust ratio, is completely negligible (Herczeg et al. 2004).  Lastly, no significant extinction is apparent in 2M1207A (as indicated by the excellent match of its SED and absolute photometry to the evolutionary models, assuming zero reddening; Fig. 14), which is located only $\sim$0.8\arcsec~ from the secondary.  We can thus confidently reject interstellar reddening as the cause of 2M1207B's anomalous behavior.      

\subsection{Edge-On Disk}
The last explanation we consider, and the simplest one consistent with our observations, is that 2M1207B suffers from {\it grey} extinction due to an edge-on disk.  In the optically thick edge-on case, the disk obscures the central object entirely, and at short (optical and near-IR) wavelengths we only see starlight scattered from the disk surfaces: a small fraction of the true luminosity of the central source.  The SED of this reflected light will closely match the intrinsic spectrum of the source (see discussion further below), yielding the correct \teff from spectral and color analyses, but the object will appear considerably sub-luminous.  The net effect is of approximately grey extinction out to near-IR wavelengths, as we observe.

Optically thick edge-on disks have now been imaged around a number of T Tauri stars (e.g., Burrows et al. 1996; Stapelfeldt et al. 1997, 1998; Padgett et al. 1999; Throop et al. 2001).  The diminution of luminosity in these cases, relative to the intrinsic brightness estimated for the central sources, is comparable to the $\sim$2.5$\pm$0.5 mag sub-luminosity we find in 2M1207B\footnote{As discussed in \S7.6, the true underluminosity of 2M1207B is likely even larger than our nominal estimate using field dwarf $BC_K$, since the latter tend to overestimate the observed \lbol for this young object.}.  For example, in the two prototypical edge-on T Tauri systems HH 30 and HK Tau/c, the observed K-band fluxes are 3--4 mags lower than expected for their age, distance and M0--M2 spectral types (Burrows et al. 1996; Stapelfeldt et al. 1998; Monin et al. 1998; Kenyon et al. 1998; this does not of course signify the true extinction in $K$ in these edge-on systems, which is extremely large, but is simply a measure of the scattered to intrinsic flux ratio).  Edge-on disks have also been proposed for some objects near and below the substellar boundary, on similar luminosity grounds (Barrado y Navascu\'{e}s et al. 2004; Luhman 2004).  Finally, the T Tauri system KH15D exhibits periodic eclipses, with depths of $\sim$3.5 mag and grey extinction of the star; this is modeled as occultation by an edge-on precessing ring or disk, with only scattered light seen during the eclipses (Herbst et al. 2002; Agol et al. 2004; Winn et al. 2004; Chiang \& Murray-Clay 2004; Knacke et al. 2004; Kusakabe et al. 2005).  These data support the viability of an edge-on disk for explaining our results.  

The disk hypothesis is also supported by analogy with the primary in the 2M1207 system.  As discussed in \S1, not only is 2M1207A already known to be surrounded by an accretion disk, but there is evidence that its disk is seen at a relatively high inclination angle, i.e., closer to edge-on than face-on, with $i$ $\gtrsim$ 60$^{\circ}$ (Scholz et al. 2005; Scholz \& Jayawardhana 2006)\footnote{Note that we do not see any underluminosity in 2M1207A, arguing against a truly edge-on configuration.  This remains compatible with the relatively high inclination suggested by Scholz et al., as discussed in \S9.}.  It does not stretch the imagination, then, to propose that the secondary has a disk as well, seen even closer to edge-on. 

Is grey extinction expected from an edge-on disk?  As a star+disk system approaches an edge-on geometry, all wavelengths are attenuated due to absorption by disk dust, but shorter wavelengths are attenuated more.  At large enough inclinations, the optical depth through the disk is too high for the shorter wavelengths to make it through at all; at this point, the star is seen at these wavelengths only in scattered light, which is minimally affected by further changes in inclination.  Longer wavelengths continue to reach us directly through the disk, albeit increasingly attenuated, to still higher inclinations, until, very close to edge-on, they too cannot pass through the disk anymore and are observed only via scattering off the disk surface.  Consequently, any given optical or infrared color first becomes redder with increasing inclination, and then reverses blueward again very close to edge-on (Whitney et al. 1997).  Detailed calculations of Class II T Tauri star+disk systems by Whitney et al. (2003) show that in almost perfectly edge-on cases, the star is seen entirely in scattered light out to near-IR wavelengths.  While the observed colors predicted by Whitney et al. for this situation are slightly bluer than the instrinsic stellar ones, due to scattering by small grains, the difference is quite small, at most a few tenths of a magnitude in the near-IR.  Modeling of brown dwarf disks by Walker et al. (2004) yields very similar results.  

These theoretical calculations imply that an optically thick edge-on disk can indeed produce nearly grey extinction into the near-IR.  Moreover, with observational errors in photometry of a few tenths of a magnitude, as in 2M1207B, any minor blueing of the intrinsic starlight by scattering off small grains would be masked, and the extinction would be indistinguishable from grey (see also \S9 below).  Observed edge-on T Tauri systems support this conclusion.  For instance, LkH$\alpha$ 263C (MBM 12A 3C), an M0 star with an almost perfectly edge-on disk, is seen only in scattered light out to the $K$-band, and the detected near-IR colors are precisely those expected intrinsically from an M0 star (Jayawardhana et al. 2002).  That is, to within the observational uncertainties in photometry, the near-IR extinction in LkH$\alpha$ 263C is wavelength-independent (i.e., grey).  

The above arguments apply to optically thick edge-on disks.  Alternatively, the disk may not be optically thick: large grains in an optically thin edge-on disk may occlude most of the starlight but allow some ($\sim$10\% in our case) to pass through.  In this case, reddening due to absorption by disk grains, and blueing due to scattering, can be negligible (producing grey extinction) if the grains are substantially larger than the observed wavelengths.  For 2M1207B, where the extinction appears roughly achromatic out to $L'$, this implies grain sizes $\gg$ 4$\mu$m.  Grey extinction due to similarly large grains has now been established in at least one T Tauri system: in the edge-on system KH15D, the constancy of the spectrum and colors in and out of eclipse out to the $K$-band, and the very weak polarization within eclipse, imply scattering by disk/ring grains with sizes $\gg$ 2$\mu$m (Agol et al. 2004; Kusakabe et al. 2005).  In addition, mid-IR photometry of our primary, 2M1207A, reveals a flat disk and negligible silicate emission from small surface grains, suggesting significant grain growth and/or settling (Sterzik et al. 2004; Gizis et al. 2005b).  It is not unlikely, then, that grains have grown large in the secondary's disk as well.  

In summary, our analysis suggests that grey extinction due to a nearly edge-disk is a viable explanation, and indeed the most plausible one, for the observed underluminosity of 2M1207B.  The information at hand is insufficient to pin down the precise disk inclination or grain sizes; however, polarization studies, currently underway, should illuminate these issues.

\section{Conclusions}

The discovery of 2M1207B, an ultra-cool and apparently planetary mass body orbiting a nearby young brown dwarf, has been tremendously exciting from a number of standpoints.  It offers us the first chance to study the atmosphere and evolution of ultra-cool, planetary-mass substellar objects in their infancy; the formation mechanism of the lowest mass brown dwarfs; and the extension of substellar binarity into the planetary mass domain.  Initial studies of 2M1207B by Chauvin et al. (2004) and Mamajek (2005), however, point to a discrepancy between its colors and approximate spectral type on the one hand, and the \teff implied by its luminosity on the other: the latter seems too low.  We have investigated this conundrum with new VLT NIR spectra and photometry of 2M1207A and B.  

Our analysis shows that 2M1207B is considerably under-luminous for its temperature, age and distance.  In particular, it appears to suffer from roughly grey extinction of $\sim$2.5 mag, in all observed bands from $\sim$$I$ to $L'$ and in $\mbol$.  It is this anomalous faintness that is responsible for the low \teff and mass derived in earlier studies.  The most plausible explanation is that 2M1207B is surrounded by a nearly edge-on disk, which occludes most (or all) of the direct light from the object and allows us to see it only in light scattered off the disk surface.  This hypothesis is compatible with observations of edge-on T Tauri systems, as well as with the known presence of a high-inclination disk around 2M1207A.  

Our edge-on disk proposal raises some questions.  First, why does a high-inclination disk lead to underluminosity in only the secondary and not in the primary?  There are two possible answers.  The first is that only a small deviation from a perfectly edge-on configuration can allow us to see the primary unocculted, if its disk has an inner hole.  For example, for a geometrically flat disk, and an inner hole $\sim$5 stellar radii in size, the primary will remain unobstructed for inclinations up to $i$ $\sim$ 80$^{\circ}$.  The primary's disk is known to be quite flat from SED analyses mentioned earlier.  Similarly, an inner hole of a few stellar radii, in an accretor such as 2M1207A, is compatible with both magnetospheric disk-accretion models as well as with the inner holes sizes inferred from the observed SEDs of other substellar accretors (Liu et al. 2003; Mohanty et al. 2004).  Since the H$\alpha$ studies of 2M1207A simply indicate $i$ $\gtrsim$ 60$^{\circ}$, and not necessarily an exactly edge-on geometry, this explanation is consistent with our proposal.  Of course, this requires that the primary and secondary disks not be exactly co-planar: edge-on in the latter, but not quite in the former.  Non-coplanar disks are known in at least some binary T Tauri systems (e.g., Monin et al. 2006), so a small difference (of order 10--20$^{\circ}$) in the disk orientations in 2M1207AB is not infeasible.  We also recall that the separation between the 2M1207 components, as mentioned in \S1, is unusually large in the context of substellar binaries.  This may point to dynamical perturbations in the past; while purely speculative, such a perturbation could misalign the component disks.  On the other hand, it may be that the 2M1207B disk is not strictly edge-on either, but is instead significantly more flared than in 2M1207A, causing occultation in the former but not in the latter for the same (high) inclination angle.  Models by Walker et al. (2004) of brown dwarf disks do predict greater flaring with decreasing sub-stellar mass (because the central gravity decreases), and we cannot rule out this possibility. 

Second, what other observational signatures do we expect from the 2M1207B disk, that might allow us to confirm its presence?  As mentioned earlier, scattering by small surface grains can cause a blueing of the starlight; conversely, at long enough wavelengths, the disk emission itself, consisting of reprocessed starlight, should begin to dominate and produce excess emission.  As discussed, disk modelings indicate that both effects can be quite small (leading to simply grey extinction) for a perfectly edge-on case, at least out to NIR wavelengths, especially if grains are large and/or in the presence of multiple scattering.  We have seen that these effects are indeed not obvious in our data: within the errors, the underluminosity in all bands is consistent with being grey.  Nevertheless, there are tantalizing hints of non-greyness at the longest and shortest bands.  In particular, the underluminosity in $JHK$ is $\sim$2.5--3 mag, while in F090M ($\sim$$I$) and $L'$ it is $\sim$2.2 mag. This is why 2M1207B appears slightly redward of the model predictions in $K$-$L'$ in Figs. 13 and 14b.  Similarly, the slight relative overbrightness in F090M, compared to $JHK$, is apparent in the small deviation in F090M from the overall predicted SED shape in Fig. 14e (this comparative overbrightness in F090M has also been noted by Song et al. 2006).  Given the uncertainties of 0.1--0.3 mag in the photometric data, these shifts are at a $\sim$2$\sigma$ level, and hence not iron-clad.  If true, however, they point to a small blueing at the shortest wavelengths and a small red excess at the longest, consistent with disk expectations.  Spatially resolved photometry of the system at still longer wavelengths, e.g. in the $M$-band, should reveal any disk excess emission more clearly.  Similarly, scattering by disk grains should produce a detectable polarization signature of at least a few percent.  These observations can confirm the presence of a disk; polarization studies can further offer insights into the grain sizes and disk orientation.  

Third, what is the disk mass we expect for 2M1207B, and is this consistent with the disk-to-star mass ratios seen in the stellar/substellar regimes?  Unfortunately, without either a spatially resolved image of the disk, or any optically thin disk-emission data (e.g., in the the mm), we cannot make any good estimates of disk mass based on our current data alone.  A very rough estimate can be acquired, however, from naive opacity arguments.  Assume a flat disk with radius $R_d$; average thickness 2$r_{\ast}$, where $r_{\ast}$ is the radius of 2M1207B (i.e., seen edge-on, the disk is just wide enough to cover the face of the central object); and average density ${\rho}_d$.  Assume further a disk grain emissivity of ${\kappa}_{\nu}$ = 0.1($\nu$/10$^{12}$ Hz)$^{\beta}$ (Beckwith et al. 1990).  Now imagine a radial line of sight through the edge-on disk towards 2M1207B.  For the disk to be optically thick along this line of sight at a given wavelength, we must have $\tau$ $\sim$ ${\rho}_d$${\kappa}_{\nu}$$R_d$ $>$ 1.  Assuming $\tau$ $\sim$ 1 then implies a lower limit on disk density, ${\rho}_d$ $\sim$ 1/(${\kappa}_{\nu}$$R_d$), and a corresponding disk dust mass of $M_{dust}$ $\sim$ 2${\pi}$$R_d$$r_{\ast}$/${\kappa}_{\nu}$.  Adopting an emissivity exponent $\beta$ $\sim$ 1 (Beckwith et al. 1990), fiducial NIR wavelength of $\sim$ 2.2$\mu$m ($K$-band), $r_{\ast}$ = 0.15\rsun (the radius predicted by the evolutionary models for a $\sim$8 \mj object at $\sim$10 Myr), and $R_d$ = 10 AU (consistent with a tidal truncation radius for the mass ratio and (projected) separation between the components) then implies $M_{dust}$ $\sim$ 7$\times$10$^{23}$ gm $\sim$ 0.01 M$_{Moon}$.  Assuming a standard gas-to-dust ratio of 100, the total disk mass is then $M_{disk}$ $\sim$ 1 M$_{Moon}$; the disk-to-star ratio is thus $\sim$ 4$\times$10$^{-6}$.  This can be compared to the average disk-to-star mass ratios, in the classical T Tauri stellar and substellar regimes, of $\sim$ 10$^{-2}$ (e.g., Scholz et al. 2006).  This calculation is obviously very simplistic; nevertheless, it qualitatively illustrates that a very small disk mass, fully admissible within the context of substellar disks, can occult 2M1207B.

A final implication of our results concerns the formation of the 2M1207AB system.  Assuming a distance of 70 pc, Chauvin et al. (2004) originally estimated a projected component separation of $\sim$55 AU, and a secondary mass of $\sim$5 M$_{Jup}$, implying a component mass ratio of $\sim$0.2.  Lodato et al. (2005) then pointed out that this separation and mass ratio are commensurate with the values observed in higher mass, {\it stellar} binaries. They also showed that the standard core-accretion mechanism of planet formation cannot produce a 5 M$_{Jup}$ secondary in this system, at the observed separation: to do so in the available time ($\lesssim$ 10 Myr, the age of the system) would require far too massive a disk around the primary (at least hundreds of M$_{Jup}$, more than an order of magnitude heavier than the primary itself).  Both facts led Lodato et al. to suggest that the 2M1207AB system probably formed in a manner akin to stellar binaries; they proposed disk fragmentation by gravitational instabilities as the most viable such mechanism.  Now, our analysis of the primary's absolute photometry shows that a distance of 53$\pm$6 pc, as derived by Mamajek (2005), is more appropriate for 2M1207AB, implying a projected component separation of $\sim$40 AU. Our inferred secondary mass of $\sim$8 M$_{Jup}$ further indicates a component mass ratio of $\sim$0.3.  Compared to Chauvin et al.'s (2004) results, our values are even more compatible with the separations and mass ratios observed in stellar binaries (see Fig. 1 of Lodato et al. 2005).  Similarly, our higher mass for 2M1207B makes its formation through core-accretion even less likely.  Our results thus strengthen the case for a binary-like origin for the 2M1207AB system, perhaps via gravitationally induced disk fragmentation.  As Lodato et al. point out, the latter mechanism still requires the primary to have started out with a disk at least as massive as the secondary, i.e., M$_{disk}$ $\gtrsim$ 0.3 M$_{primary}$.  This is an order of magnitude higher than the disk masses currently inferred, from sub-mm/mm fluxes, for brown dwarfs that are $\sim$1--few Myrs old (M$_{disk}$/M$_{BD}$ $\sim$ few \%; Klein et al. 2003; Scholz et al. 2006).  However, disk masses do decline with time; while stellar T Tauri disks, at a few Myrs, also contain only a few percent of the mass of the central star, disks around younger embedded protostars comprise a large fraction of the central mass.  Hence a relatively massive disk around 2M1207A much earlier in its life, when fragmentation is believed to occur, is not entirely implausible.  


Lastly, it is important to remember that, while our mass for 2M1207B exceeds the estimates of Chauvin et al. (2004) and Mamajek (2005), it is still in the planetary regime.  It is fascinating to imagine that stellar binary formation mechanisms, when translated to the low mass substellar domain, may directly form planetary mass companions to brown dwarfs.  Our data moreover indicate that such companions can even possess disks of their own; perhaps, we speculate, capable of forming planetesimals, moons and asteroids. Further study of 2M1207AB promises rich insights into substellar formation, evolution and properties.  
      
\acknowledgments
We would like to express our gratitude to the staff of ESO-VLT, and especially to Nancy Ageorges for her technical support and advice.  S.M. would like to sincerely thank Gael Chauvin for advice on data reduction and very helpful inputs on the results, Frank Marchis for an independent deconvolution of our $J$-band images and a check on our photometry, Adam Burgasser and Gilles Chabrier for their suggestions and close examination of the analysis, and Kevin Luhman and Gibor Basri for many illuminating discussions that shaped this paper.  We are particularly indebted to Derek Homeier, France Allard, Isabelle Baraffe, Peter Hauschildt and Didier Saumon for providing exhaustive synthetic spectra and evolutionary models at all hours of day and night, and invaluable input on atmospheric behavior.  We also thank Adam Burgasser for an extremely helpful and detailed referee report.  S.M. is grateful as well to the Spitzer Fellowship program for funding this research; the work was also supported in part by University of Toronto start-up funds to R.J.  E.M. is supported through a Clay Postdoctoral Fellowship from the Smithsonian Astrophysical Observatory.

\clearpage

\plotone{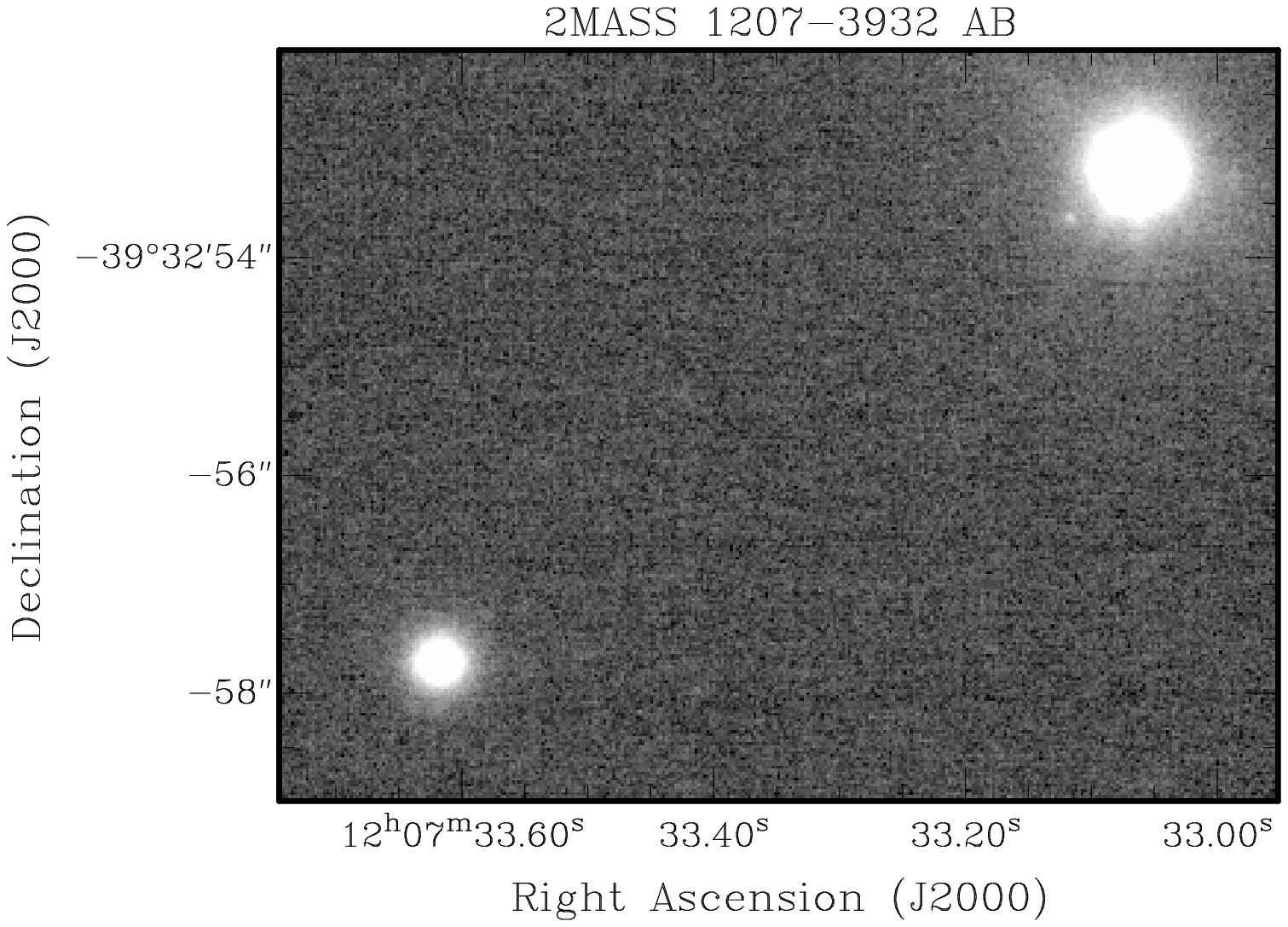}
\figcaption{\label{fig1}Our final reduced $J$-band image, showing the 2M1207AB system in the upper right (the secondary is visible near the lower left rim of the primary), and a background K-type giant in the lower left corner.  North is up, East is left; the separation between 2M1207A and the K giant is $\sim$7$\arcsec$. }

\plotone{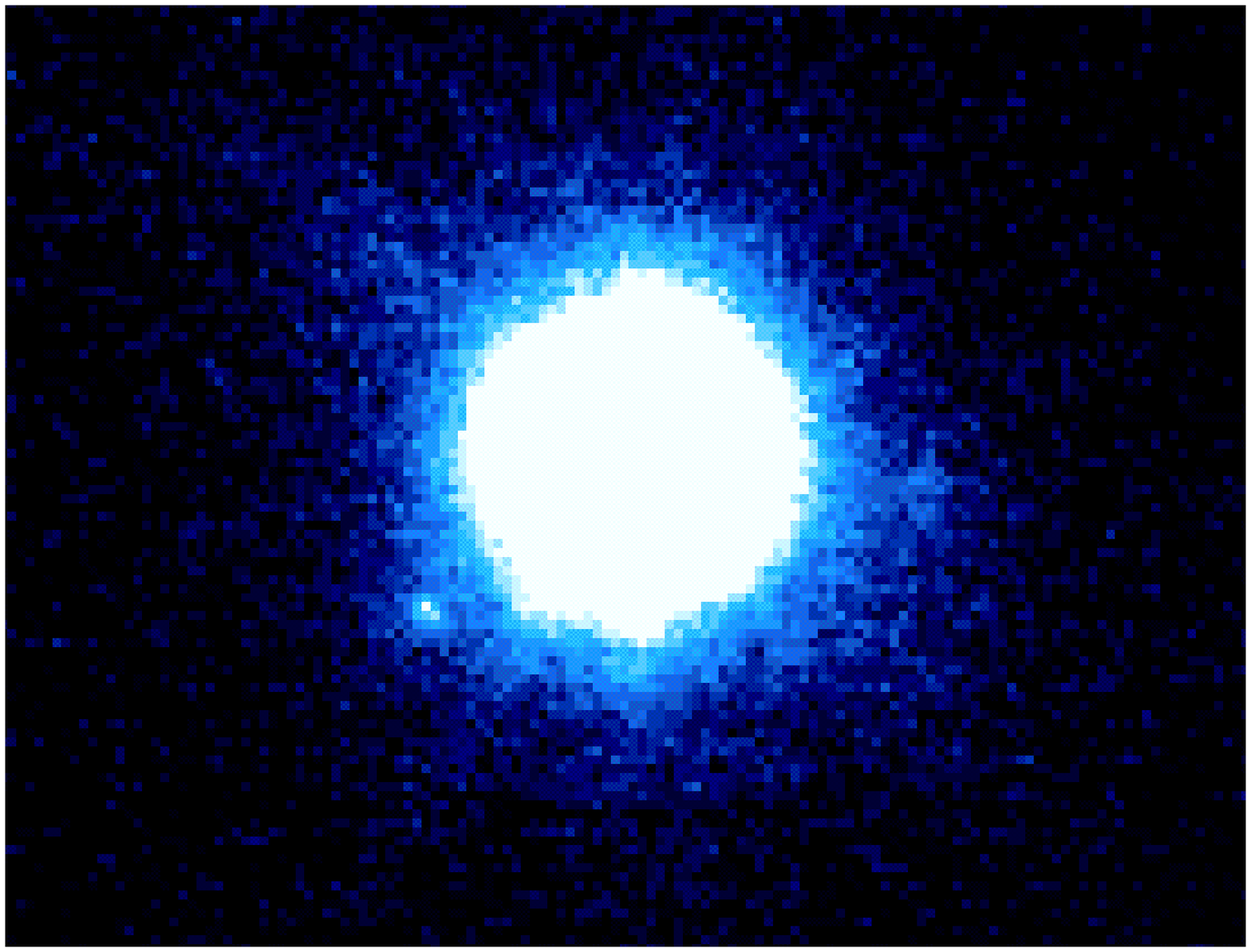}
\figcaption{\label{fig2} Zoom-in view of Fig.1.  North is up, East is left.  2M1207A is the central bright source, 2M1207B the bright dot to its immediate lower left, at a separation of 769$\pm$10 mas and position angle (E of N) of 125.6$\pm$0.7$^{\circ}$.  }

\plotone{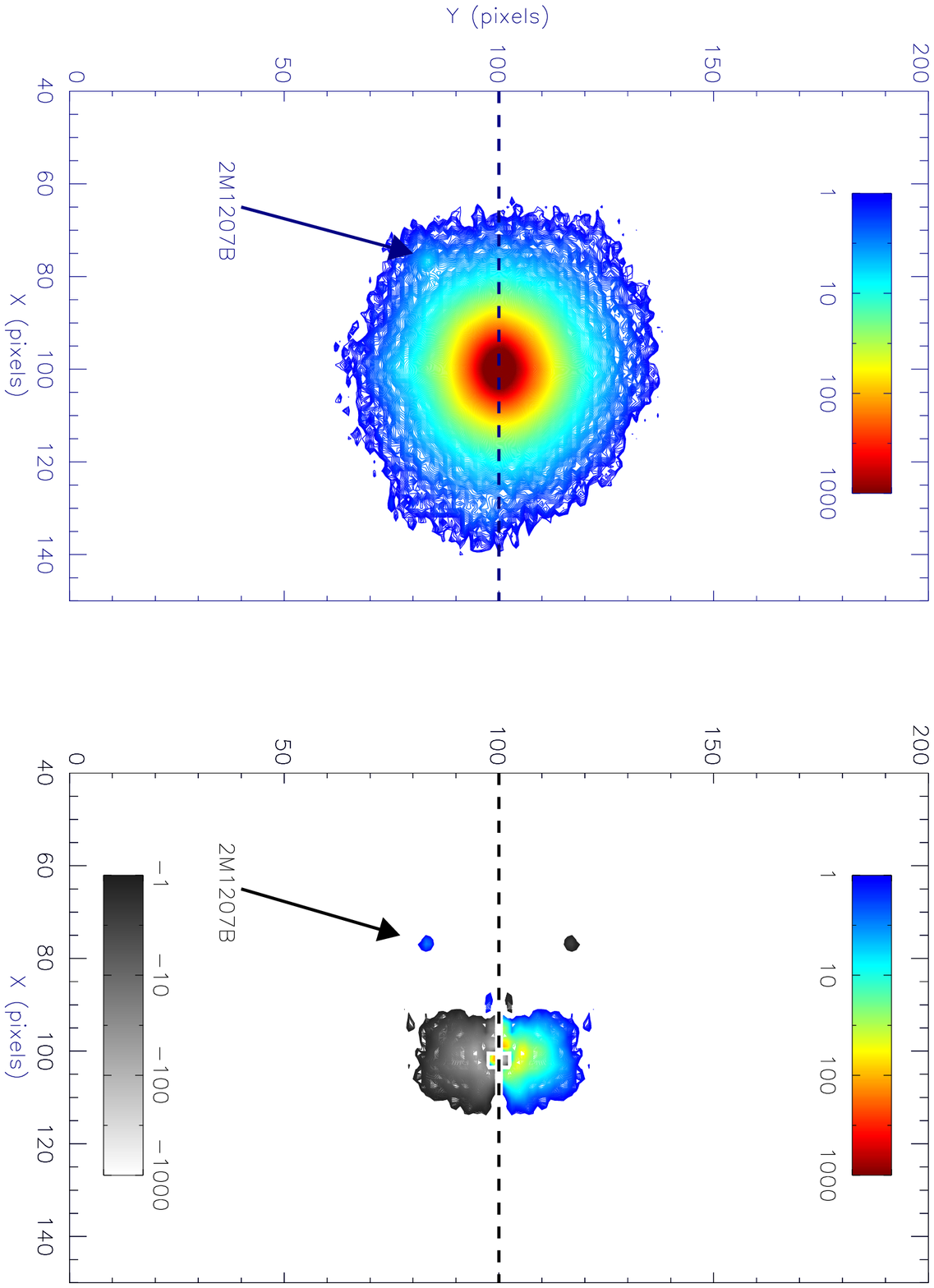}
\figcaption{\label{fig3} Contour plot illustration of our $J$-band photometry reduction procedure.  {\it Left}: Contour plot of the combined PSFs of 2M1207A and B, with observed counts (from +1 to +1000) in logarithmic units.  The secondary, marked by an arrow, is visible as a faint source sitting on the wings of the primary.  The primary's PSF is highly symmetric about the $X$-axis drawn through its centroid ({\it thick dashed line}).  To extract the secondary, we reflect the primary's PSF about this axis, and subtract from the original image.  {\it Right}: Contour plot of the residuals after subtraction, in logarithmic units.  Positive residuals (+1 to +1000) are shown in color, on the same scale as in the left panel; negative (-1 to -1000) in greyscale.  The residuals on either side of the $X$-axis are +ve and -ve mirror images.  
The procedure yields a clean image of the secondary (marked by arrow); residuals from the primary are $<$10\% of the original counts everywhere. }

\plotone{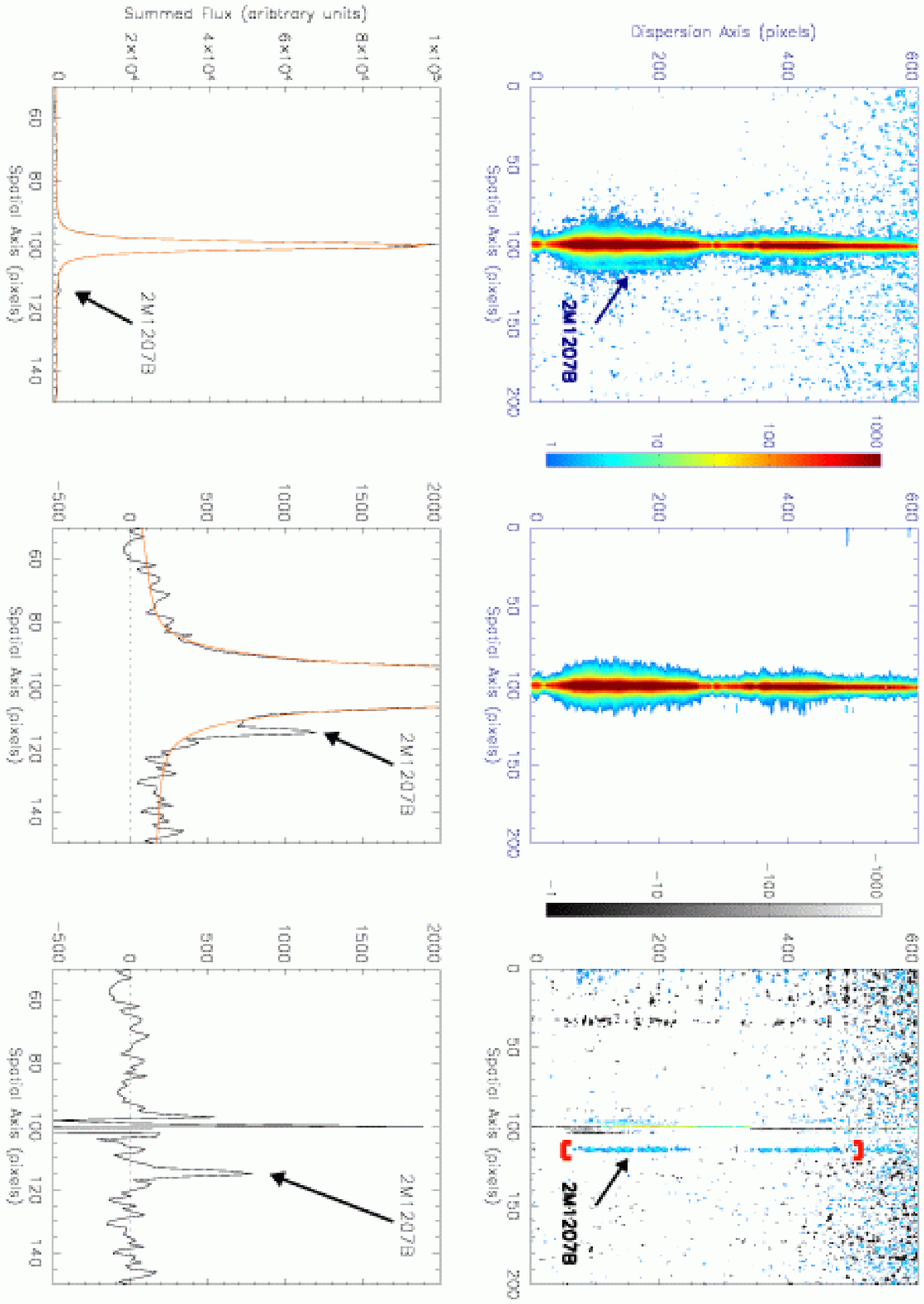}
\figcaption{\label{fig4} Illustration of our $HK$ spectral reduction procedure, for a representative OB (from April 4 2005). {\bf Top panels}: {\it Left}: Contour plot of our original smoothed median 2-D spectral image for this OB, with counts (from +1 to +1000) in logarithmic units.  The central bright spectrum is that of the primary; the secondary is the faint strip on the right wing of the primary (marked by arrow).  Wavelength increases along $y$-axis, from bottom to top, with the $H$-band in the bottom half and $K$-band in the top half of the image.   {\it Middle}: Our model 2-D spectrum for the primary (see \S4.2), on the same scale.  {\it Right}: Residuals, in logarithmic units, after subtracting model from original.  Positive residuals (+1 to +1000) shown in color, on the same scale as in the other two panels; negative residuals (-1 to -1000) in greyscale.  
The procedure efficiently removes the primary (residuals everywhere are $<$10\% of the original counts), leaving behind a clear spectrum of the secondary (marked by arrow).  The red brackets indicate the region of the secondary's spectrum we actually use in our subsequent analysis; the reddest parts of the $K$-band are noisy, and thus excluded.  {\bf Bottom panels}:  Alternative representation of the same reduction.  {\it Left}: Spatial profile of our original 2-D spectral image, after summing over dispersion axis.  Central peak is 2M1207A; secondary marked by arrow.  {\it Middle}: Zoom-in of model spatial profile ($red$) overplotted on the data ($black$), showing an excellent fit to the primary, including in the vicinity of the secondary.  {\it Right}: Residual spatial profile after model subtraction, showing a clean profile of 2M1207B alone; primary residuals are less than $<$10\% of the original counts everywhere, and insignificant near the secondary.}

\plotone{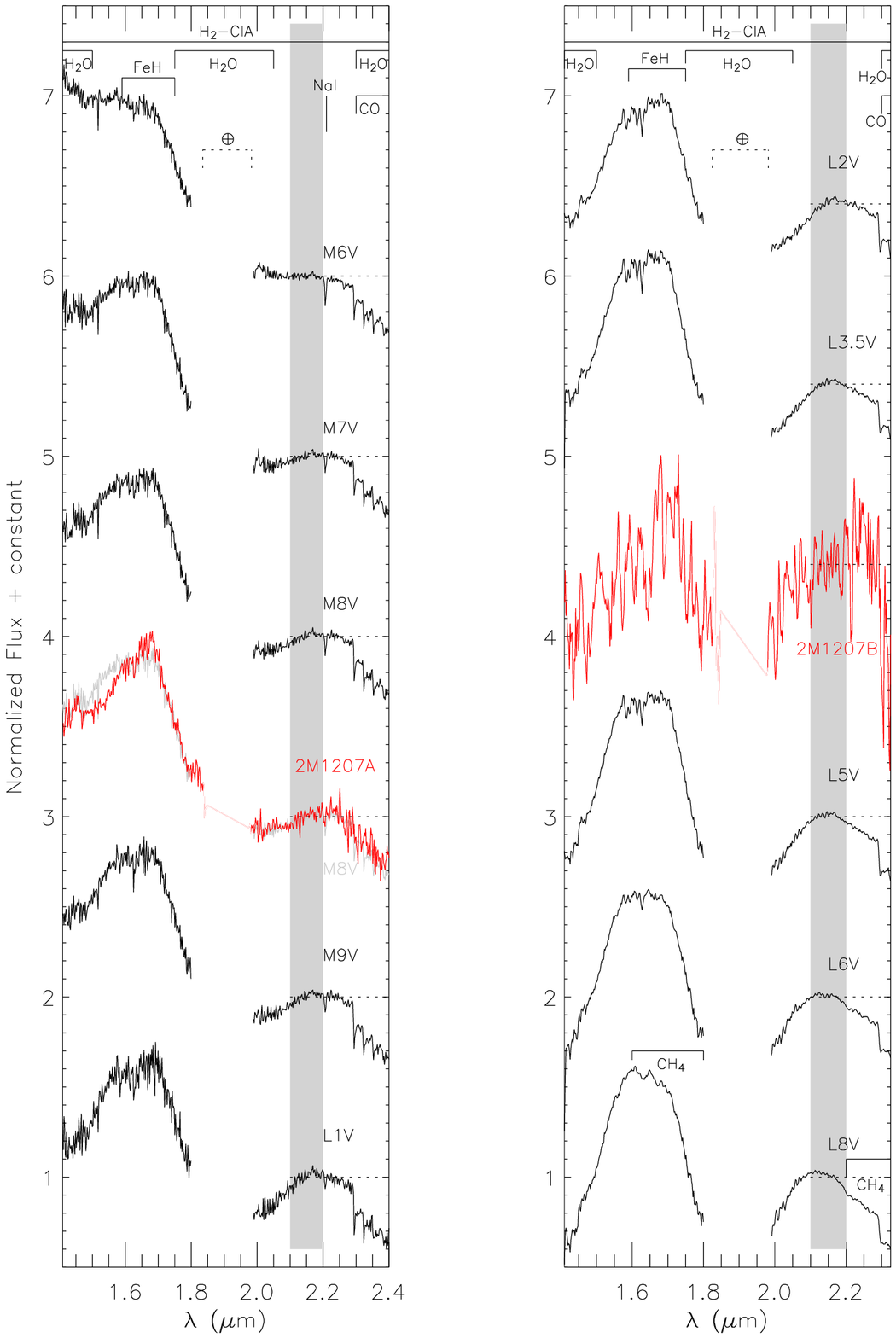}
\figcaption{\label{fig5} Comparison of our observed $HK$ spectra of 2M1207A ($red$, {\it left panel}) and 2M1207B ($red$, {\it right panel}) to the spectra of field M and L dwarfs ({$black$}).  All spectra have been normalized by their mean flux over 2.1--2.2$\mu$m (region marked by grey strip), and offset for clarity (normalization levels shown by dotted lines).  Opacity sources in various wavelength regions are marked; the 1.8--2.0$\mu$m interval is dominated by telluric absorption, and has been masked.  The primary spectrum is overall similar to that of field M8--M9 dwarfs (M8 dwarf overplotted in $grey$ for comparison), but shows a sharply peaked $H$-band profile and negligible \na absorption, indicating low gravity.  In the secondary, the overall strength of the \h2o absorption in $H$ and $K$ is similar to that in mid-to-late L dwarfs.  However, the $H$-band profile is sharply peaked like in the primary, and the spectrum from $H$ to $K$ is much redder than in the field dwarfs, indicating low gravity and a dusty atmosphere.  See \S6.2.2.}

\plotone{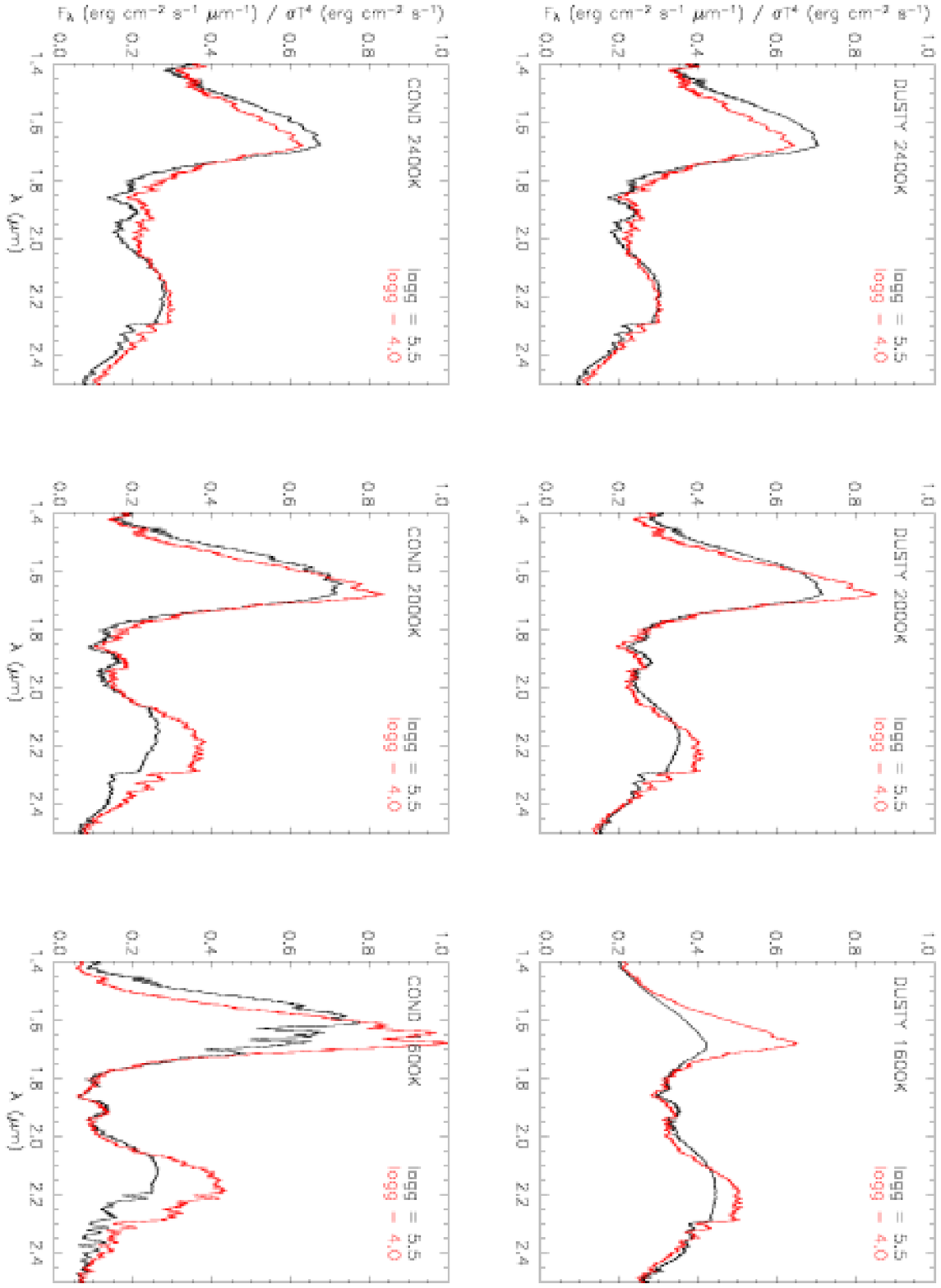}
\figcaption{\label{fig6} DUSTY and COND model fluxes ($top$ and $bottom$ panels respectively), at different gravities (log $g$=4.0 in $red$ and 5.5 in $black$) and \teff (2400, 2000 and 1600K, $left$ to $right$).  Spectra at every \teff are normalized by the corresponding bolometric flux $\sigma$T$_{eff}^4$, to clearly illustrate relative differences due to various gravity, \teff and dust related opacity effects. See \S7.1.1.}

\plotone{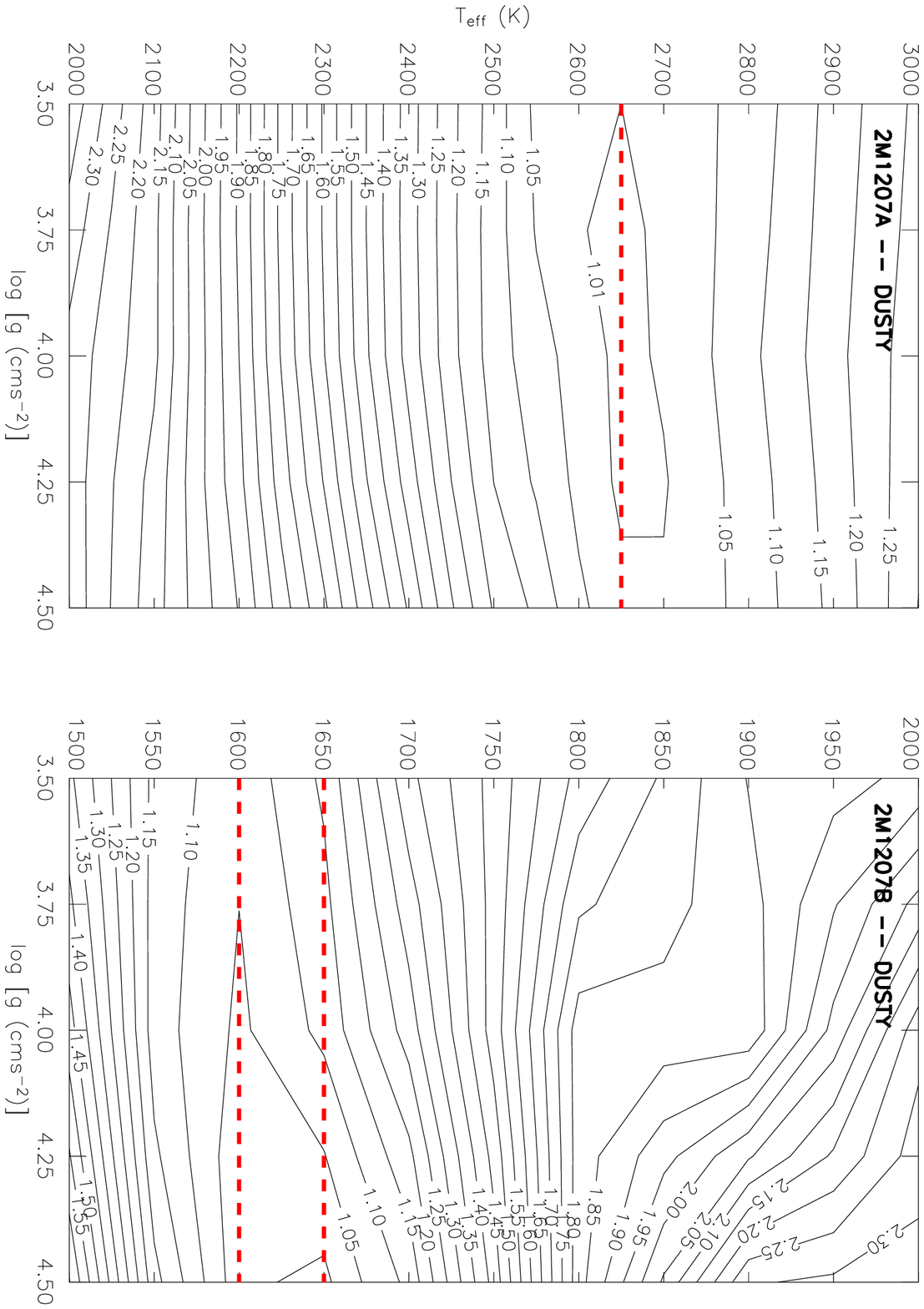}
\figcaption{\label{fig7} Contour plots of least-squares values ($s^2$) for DUSTY model fits to 2M1207A and B.  {\it Fig. 7a (left)}: Contours for 2M1207A.  All $s^2$  normalized by the minimum (best-fit) value, obtained at [\teff, log $g$] = [2600K, 4.0], to show relative merit of the fits.  The fits are relatively insensitive to gravity, with best-fit \teff $\approx$ 2600K over log $g$ = 3.5--4.5 (marked by $red$ dotted line).  Note that the merit of the fits changes relatively slowly for \teff $\gtrsim$ 2700K (contours are widely spaced) compared to below 2700K (i.e., upper \teff limit not as well constrained as lower).  {\it Fig. 7b (right)}: Contours for 2M1207B.  All $s^2$ normalized by minimum (best-fit) value, obtained at [\teff, log $g$] = [1650K, 4.5].  The fits are again relatively insensitive to gravity, with best-fit \teff $\approx$ 1600K over log $g$ = 3.5--4.0 and 1650K at log $g$ = 4.5 (marked by $red$ dotted lines).  See \S7.1.3. }

\plotone{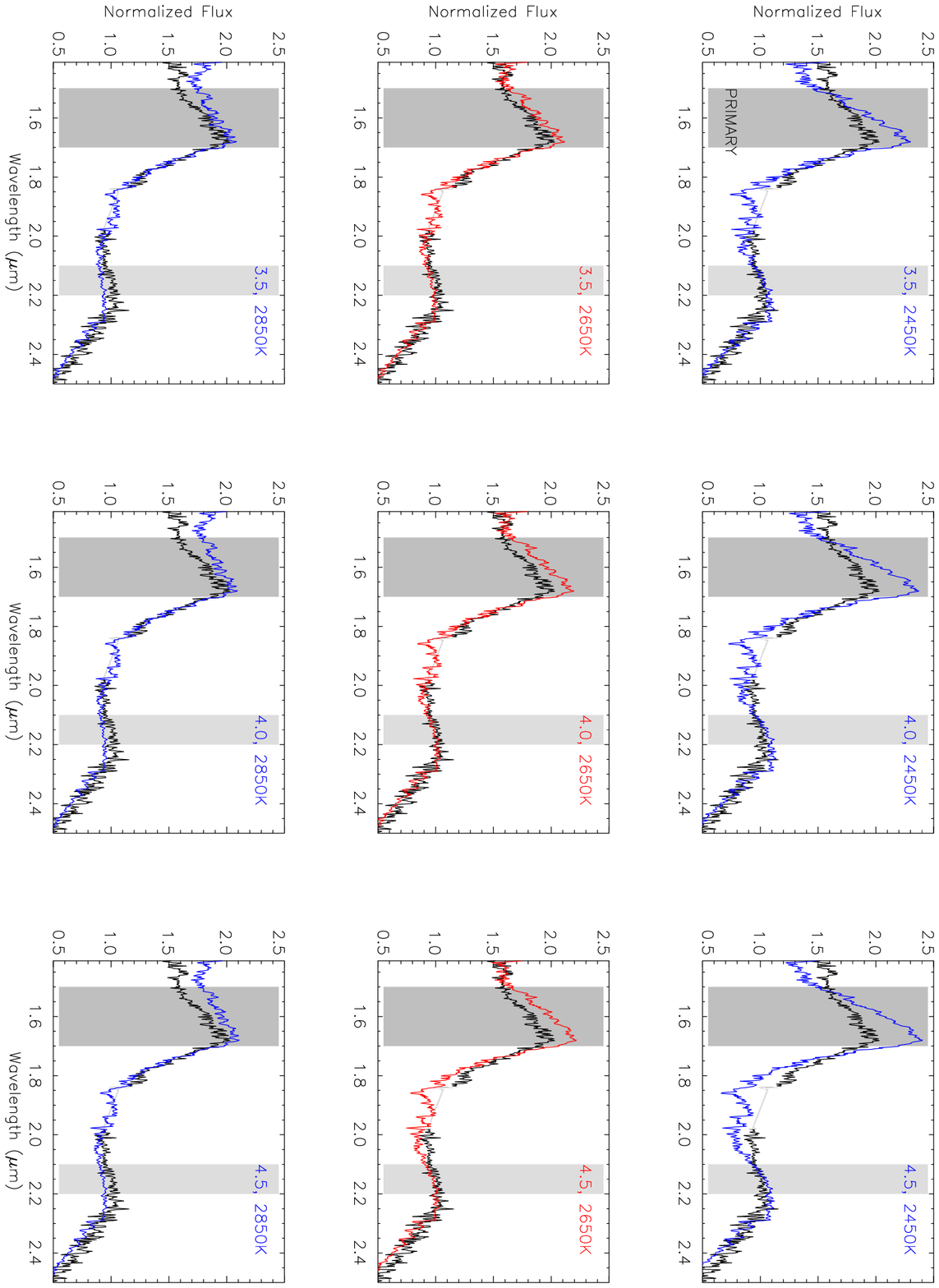}
\figcaption{\label{fig8} Observed $HK$ spectrum of 2M1207A ($black$) compared to DUSTY models at various \teff ($top$ to $bottom$) over the expected range in gravities (log $g$ = 3.5--4.5, {\it left} to {\it right}.).  {\it Dark grey} strip on left of each plot shows region excluded during fitting (1.5--1.7$\mu$m).  The observed spectrum is normalized by its mean flux over 2.1--2.2$\mu$m (region marked by {\it light grey} strip on right of every plot).  Each model is shown at its optimum normalization, as described in \S7.1.2.  Best-fit models (see contour plot in Fig. 7a), at 2650K, shown in {\it red}; acceptable range in \teff fits (2650$\pm$200K) shown in $blue$.  Note the overluminosity in the models over 1.5--1.7$\mu$m compared to the data at all \teff and log $g$ shown, due to model \h2o and FeH opacity problems; this is why this region is excluded in the fits.  See \S7.1.3.  }

\plotone{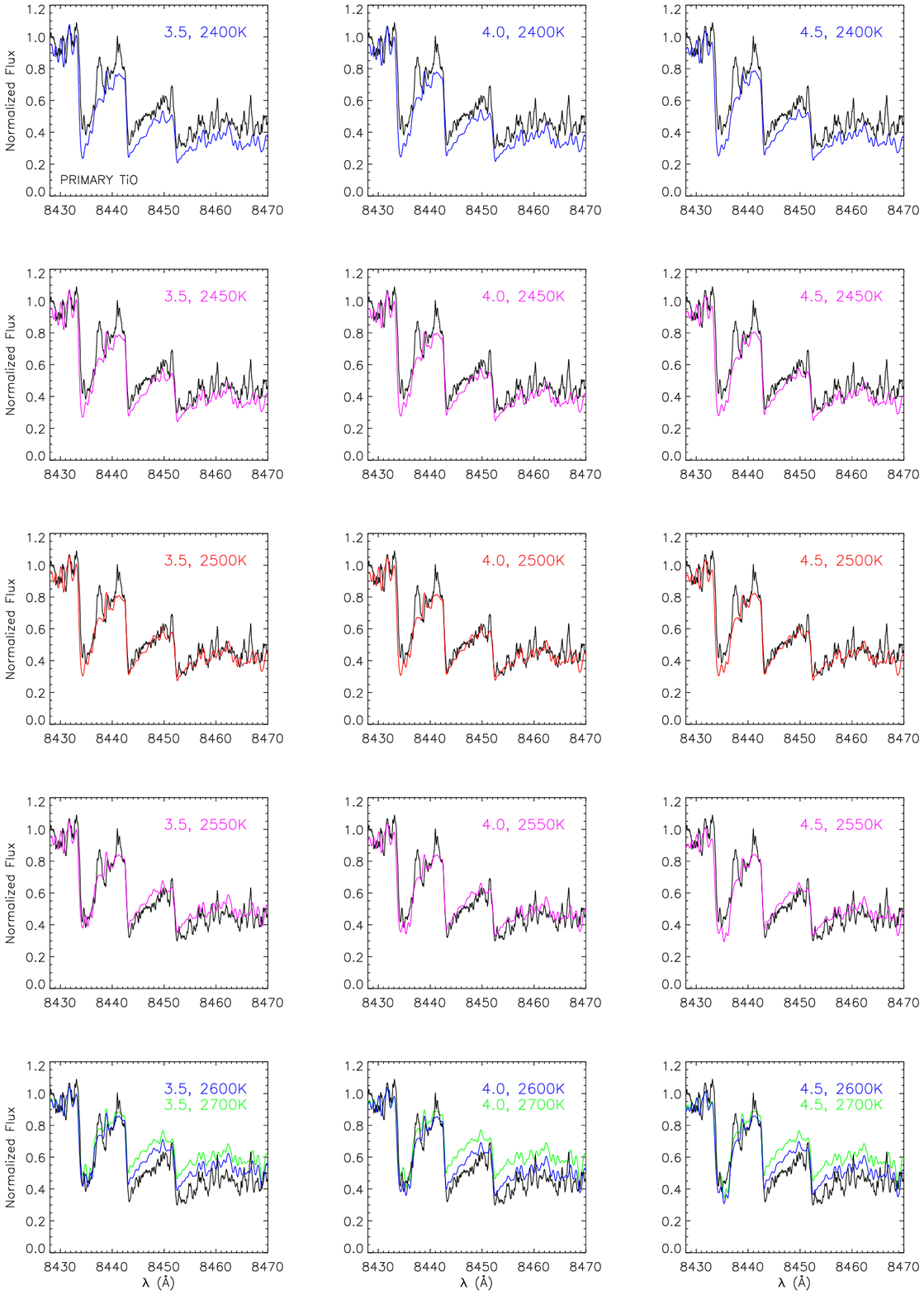}
\figcaption{\label{fig9} Observed TiO triple-headed band in high-resolution optical spectrum of 2M1207A ($black$), compared to DUSTY models at various \teff ($top$ to $bottom$) over the expected range in gravities (log $g$ = 3.5--4.5, from $left$ to $right$).  Best-fit models, at 2500K, shown in $red$; models offset from this by $\pm$50K shown in $magenta$; models offset by $\pm$100K shown in {\it blue}.  The TiO bands are highly insensitive to gravity but extremely sensitive to \teff: offsets of $\pm$50K from the best-fit produce a small but discernible worsening in the fit, and offsets of $\pm$100K are clearly worse.  Note that the 2700K models ($green$, bottom row) are completely inconsistent with the TiO data.  See \S7.1.3. }

\plotone{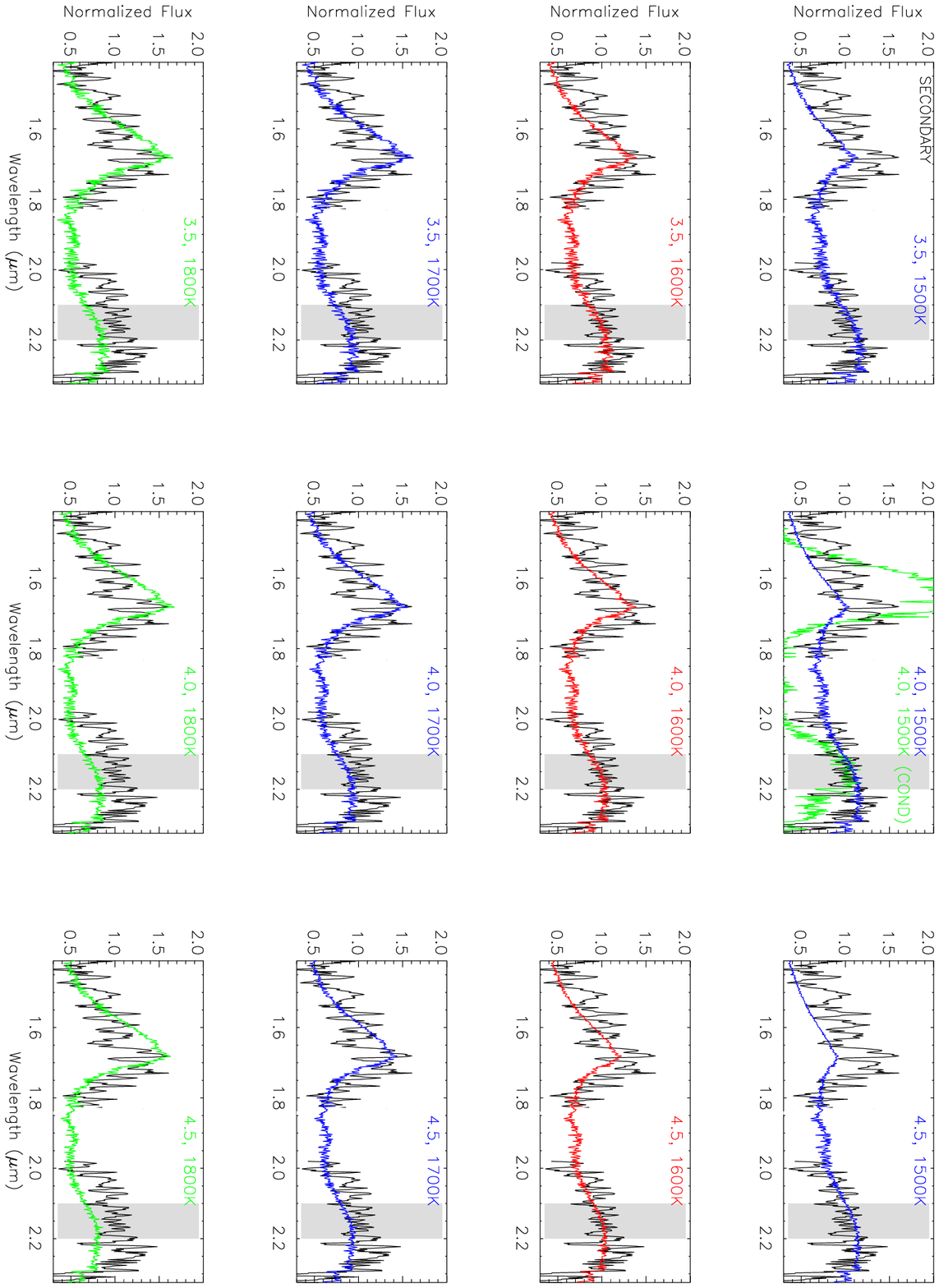}
\figcaption{\label{fig10} Observed $HK$ spectrum of 2M1207A ($black$) compared to DUSTY models at various \teff ($top$ to $bottom$) over the expected range in gravities (log $g$ = 3.5--4.5, {\it left} to {\it right}.).  Data and models normalized in same fashion as in Fig. 8.  Best-fit models (see contour plot in Fig. 7b), at 1600K, shown in {\it red}; acceptable range in \teff fits (1600$\pm$100K) shown in $blue$; models at 1800K shown in {\it green} for comparison.  The best-fit models reproduce the data well, within the uncertainties of the noise.  The DUSTY fit is clearly very poor by 1800K.  A COND model at 1500K and log $g$=4.0 ($green$, {\it top middle} panel) is also shown for comparison, and is a very poor fit to the data: it is much bluer than 2M1207B, and also shows strong CH$_4$ absorption at $\geq$2.2$\mu$m, which is entirely absent in the data. See \S7.1.3. }

\plotone{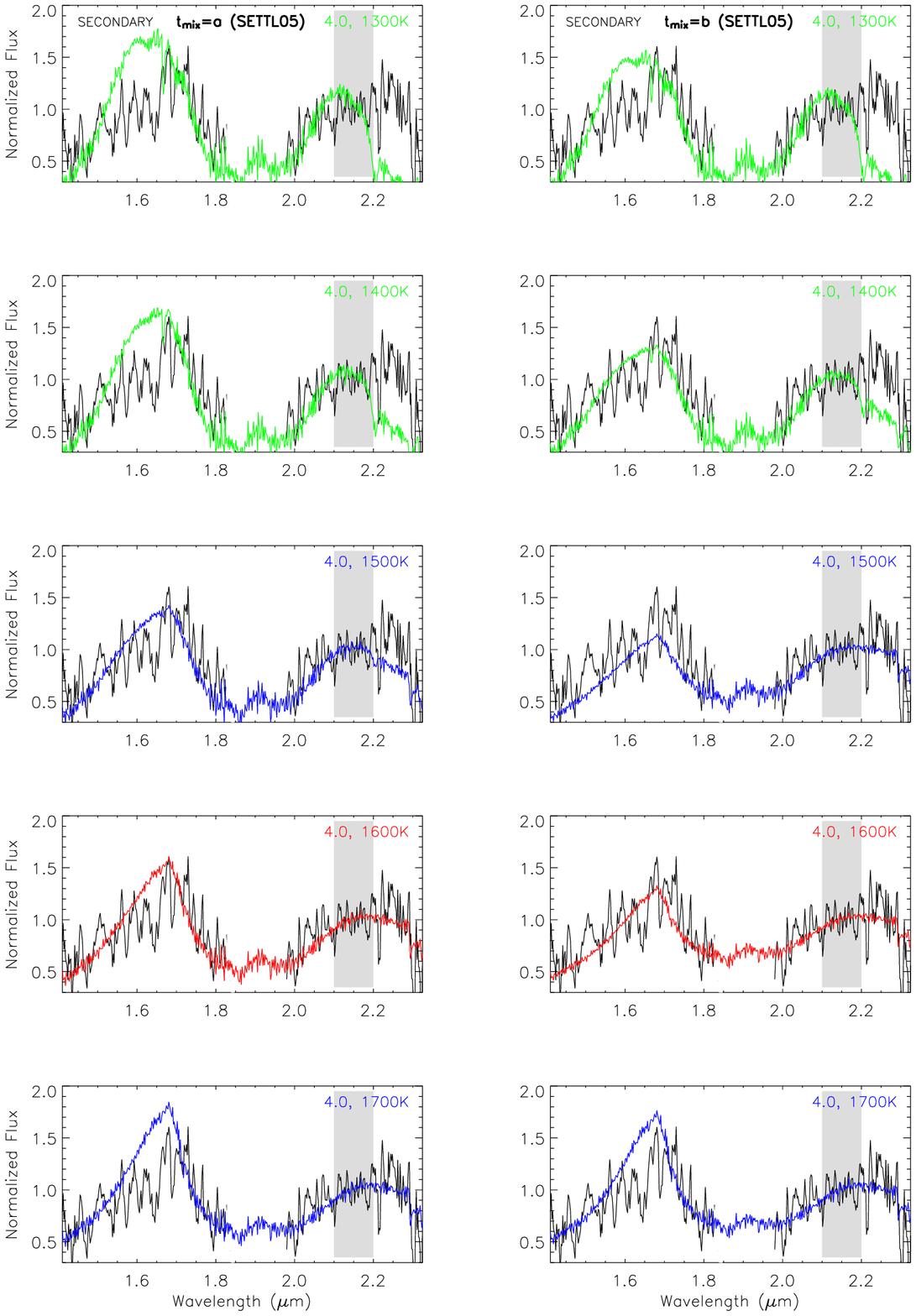}
\figcaption{\label{fig11} Observed $HK$ spectrum of 2M1207B ($black$) compared to SETTL models at various \teff ($top$ to $bottom$), at the representative gravity of log $g$ = 4.0.  Data and models normalized in same fashion as in Fig. 8.  $Left$ panels show SETTL models with less efficient mixing (more settling), compared to the models on the $right$ (see model description in \S5).  For both sets of models, the best-fit \teff, 1600K, is shown in red; the range in \teff fits bracketing the data (1600$\pm$100K) shown in $blue$; models with \teff $<$ 1500K (1400 and 1300K) shown in $green$.  The best-fit SETTL models reproduce the data quite well, similar to the best-fit DUSTY models in Fig. 10; the `less settling' model, on the right, appears to be a slightly better match to the observed $H$-band profile.  Models at $<$ 1500K are very poor fits to the data; at these \teff, much of the dust has settled out of the atmosphere in these models, bringing them closer to COND: they are much bluer than the observed spectrum, with strong CH$_4$ absorption at $\geq$2.2$\mu$m that is not evident in the data. See \S7.1.3.  }

\plotone{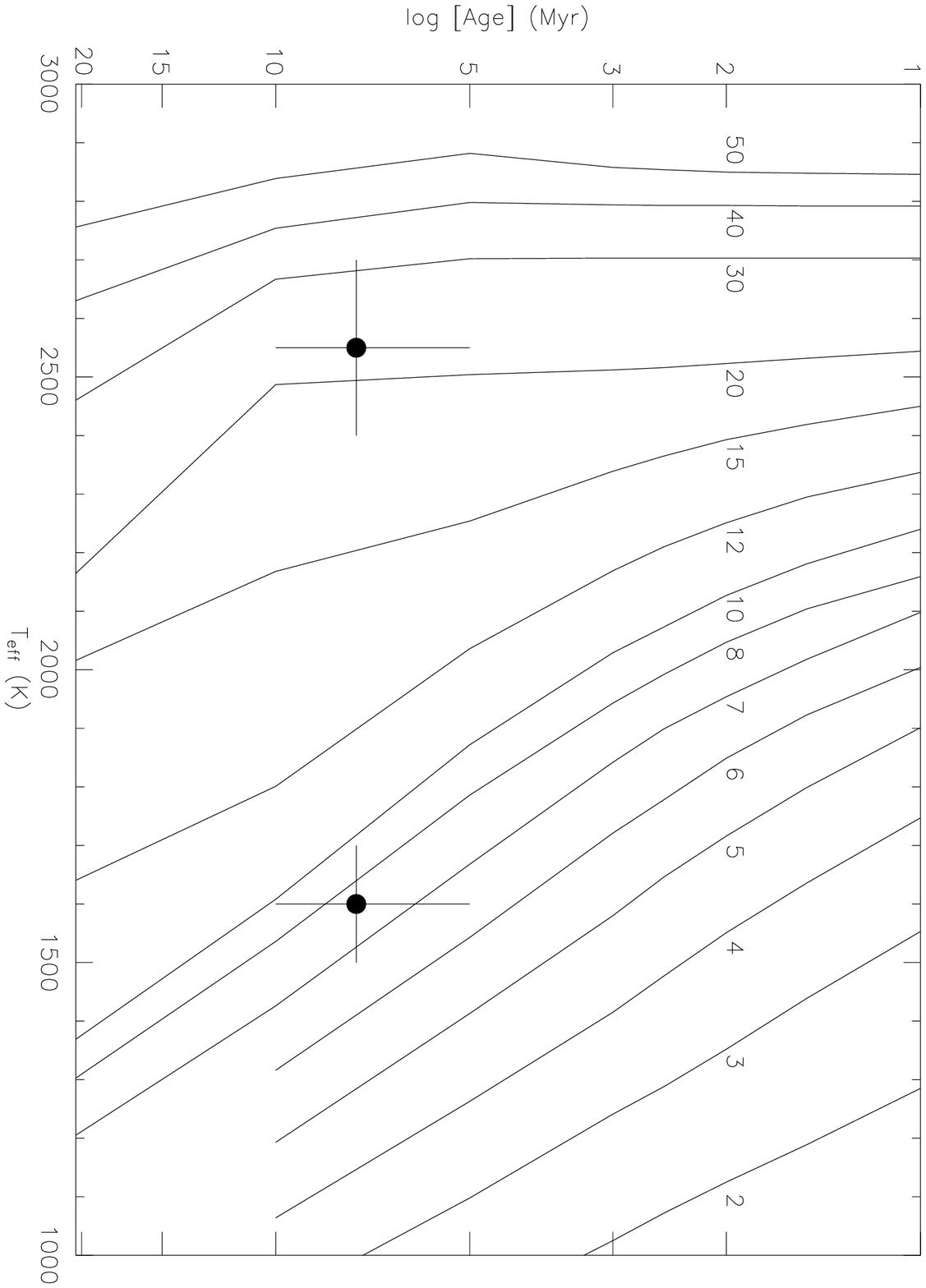}
\figcaption{\label{fig12} Comparison of 2M1207A and B to the Lyon theoretical evolutionary tracks, in the age-\teff plane.  For our adopted age (5--10 Myr), and spectroscopically derived \teff for the two components (primary: 2550$\pm$150K, secondary: 1600$\pm$100K), the tracks imply masses of $\sim$18--30 \mj and 6--10 \mj for 2M1207A and B respectively.  See \S7.2.} 

\plotone{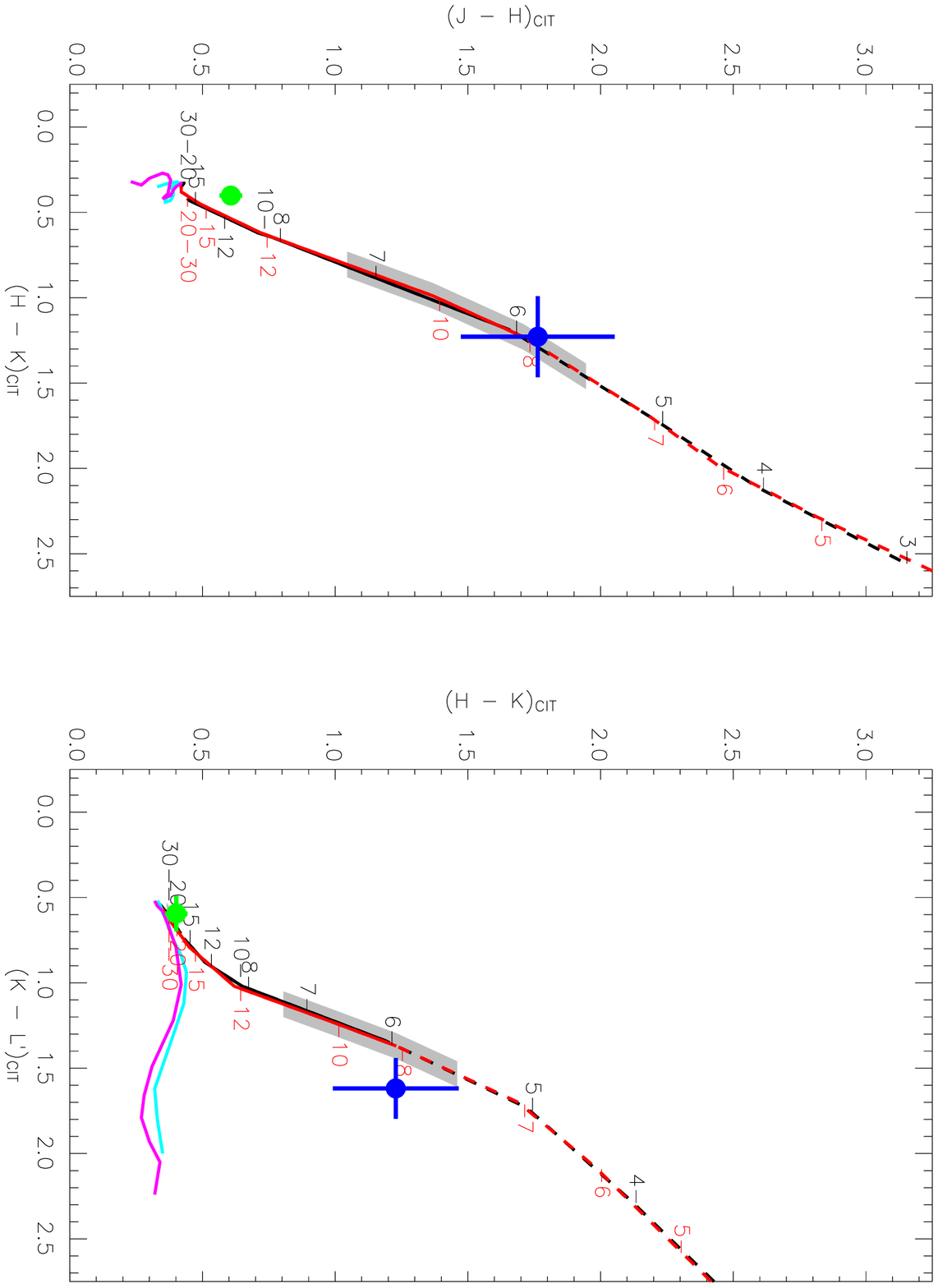}
\figcaption{\label{fig13} Comparison of 2M1207A and B to Lyon theoretical color-color diagrams.  All observed $JHK$ converted to CIT system (see \S5).  DUSTY isochrones at 5 and 10 Myrs shown in $black$ and $red$ respectively, with the corresponding masses marked; COND isochrones at 5 and 10 Myrs for the same range in masses shown in $aqua$ and $magenta$ respectively.  {\it Green filled circles} represent the observed colors of 2M1207A, and {\it blue filled circles} denote 2M1207B; 1$\sigma$ error bars in photometry shown for both.  The {\it grey filled zones} indicate the range in model colors corresponding to our spectroscopically derived range in \teff (1600$\pm$100K) for 2M1207B.  {\it Left}: $J$-$H$ vs. $H$-$K$.  {\it Right}: $H$-$K$ vs. $K$-$L'$. See \S7.3. }

\plotone{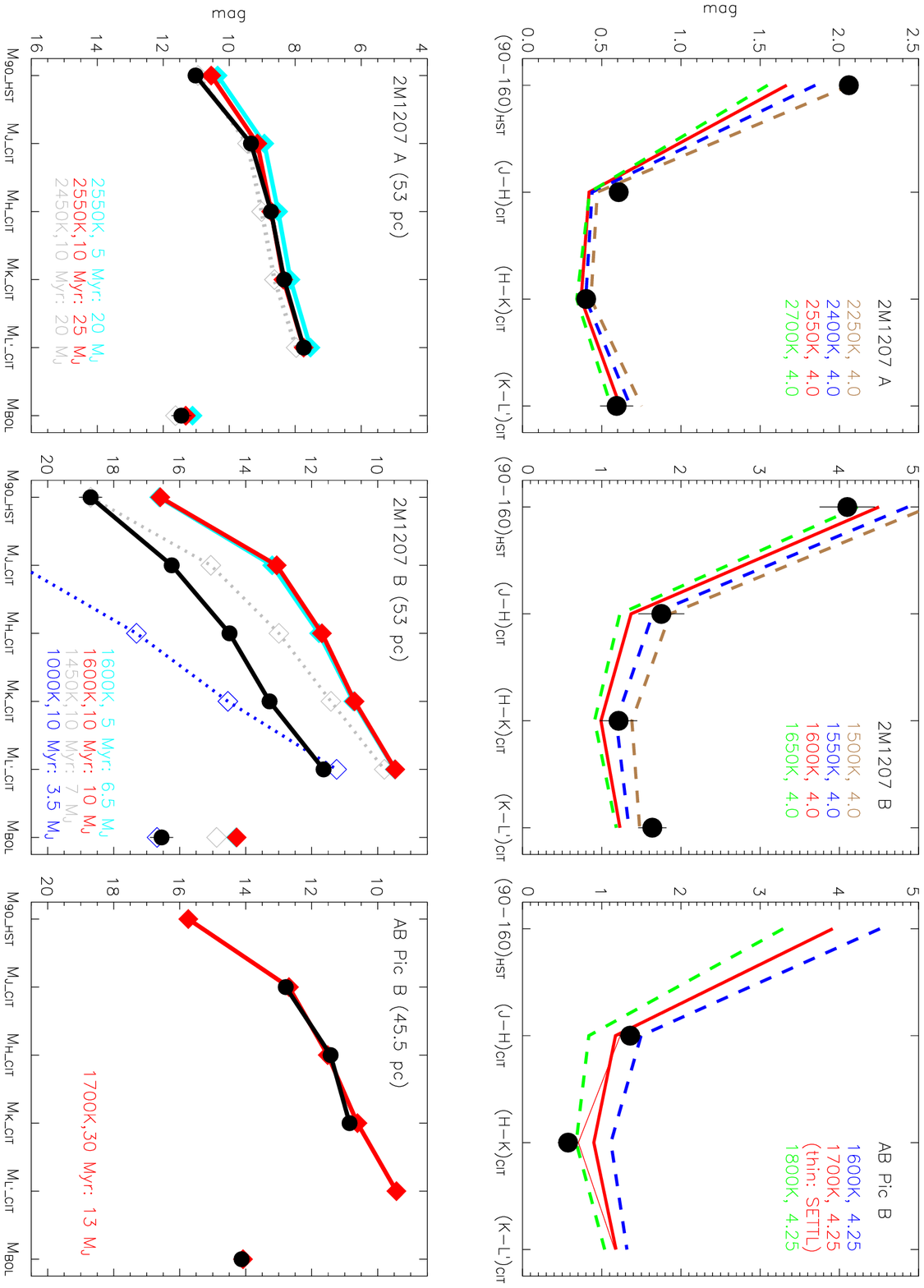}
\figcaption{\label{fig14} {\bf Top panels}: Comparison of the observed colors of 2M1207A, 2M1207B and AB Pic B to model atmosphere predictions.  All models shown are DUSTY, unless otherwise noted.  {\it Fig. 14a (top left)}: 2M1207A. {\it Black filled circles} are the data: HST [F090M-F160W] ($\sim$ $I$-$H$) and ground-based $J$-$H$, $H$-$K$ and $K$-$L'$, from left to right.  Models at various \teff for log $g$ = 4.0 are shown as colored lines (see key in plot).  
{\it Fig. 14b (top middle)}: Same, for 2M1207B.  The $y$-axis scaling here is twice that for 2M1207A.  
{\it Fig. 14c (top right)}: Same, for AB Pic B (but with only the available $J$-$H$ and $H$-$K$ data shown).  Models are now for log $g$ = 4.25.  The {\it thick} lines show DUSTY models at 1600, 1700 and 1800K, while the {\it thin red} line shows a SETTL model at 1700K.  
{\bf Bottom panels}: Comparison of the SEDs and $\mbol$ of 2M1207A, 2M1207B and AB Pic B to the evolutionary model predictions.  All models shown use DUSTY atmospheres.  The observed photometry has been converted to absolute magnitudes using $d$ = 53 pc for 2M1207A and B, and $d$ = 45.5 pc for AB Pic B.  {\it Fig. 14d (bottom left)}: 2M1207A.  Black circles show the data: M$_{F090M}$, M$_J$, M$_H$, M$_K$, M$_L'$ and $\mbol$ from left to right.  Model predictions are shown for \teff = 2550K at 10 Myr ($red$) and 5 Myr ($aqua$), and \teff = 2450K at 10 Myr ($grey$ $dashed$) .  
{\it Fig. 14e (bottom middle)}: Same, for 2M1207B.  Same $y$-axis scaling as the left panel.  Model predictions are shown at \teff = 1600K at 10 Myr ($red$) and 5 Myr ($aqua$); \teff = 1050K at 10 Myr ($blue$ $dashed$); and \teff = 1450K at 10 Myr ($grey$ $dashed$).    
{\it Fig. 14f (bottom right)}: Same, for AB Pic B (but with only the available $J$, $H$ and $K$ data shown).  The model prediction is shown for \teff = 1700K at 30 Myr ($red$).   
See \S7.4--7.6.  }   

\plotone{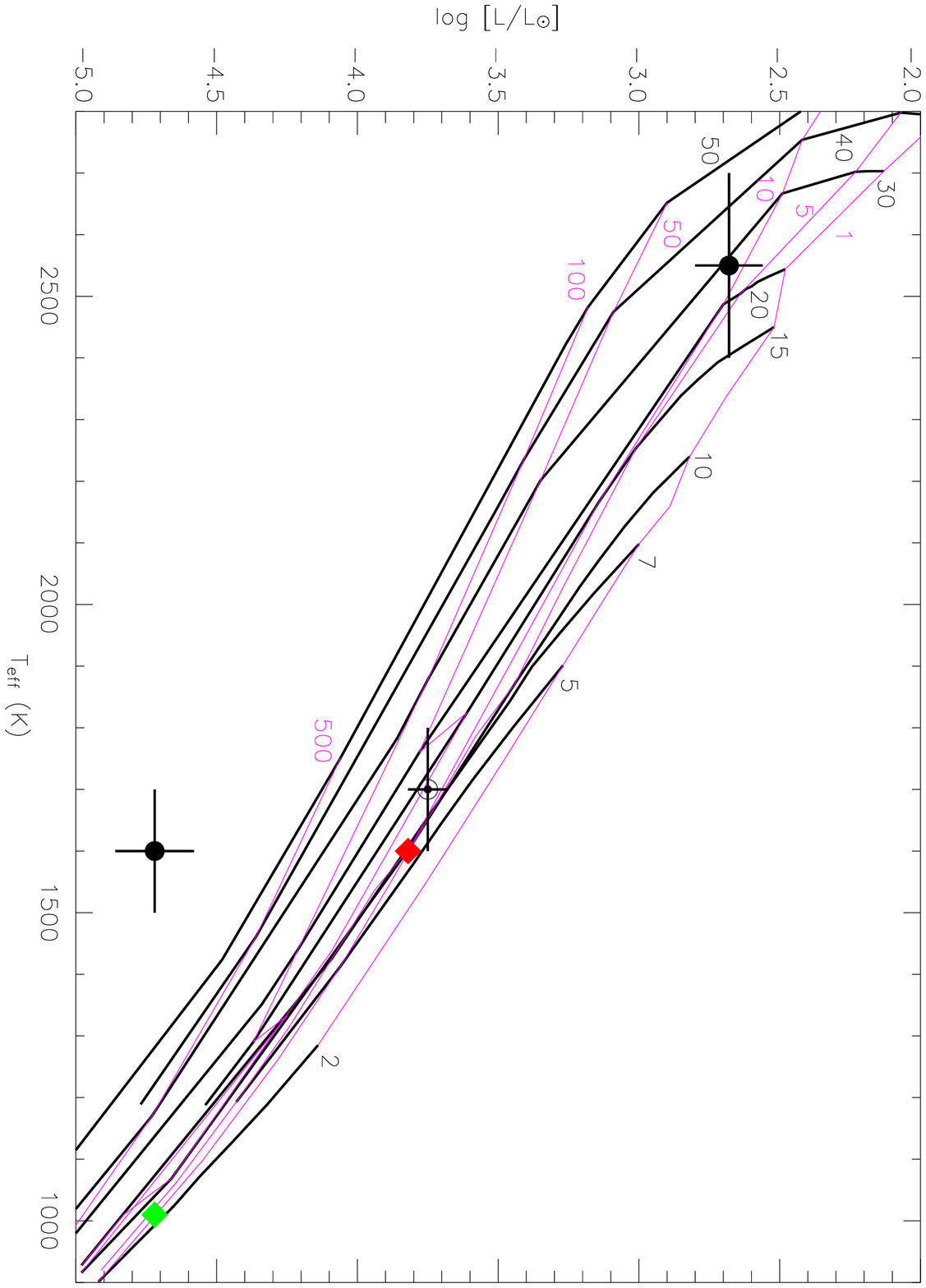}
\figcaption{\label{fig15} Comparison of the derived \teff and \lbol for 2M1207A, 2M1207B and AB Pic B to the Lyon theoretical HR-diagram.  {\it Black lines} denote evolutionary tracks for masses of 2--50 \mj (masses marked on plot).  {\it Magenta lines} denote isochrones at 1, 5, 10, 50, 100 and 500 Myr (marked on plot; age increases from top to bottom).  The observed position of 2M1207A is completely consistent with a mass of 20--30 \mj and age of 10 Myr, in agreement with our previous results.  Similarly, for its derived \teff and \lbol, AB Pic B ({\it bullseye}) lies at its expected age of $\sim$30 Myr.  The \teff and \lbol of 2M1207B, however, imply an age $>$ 0.5 Gyr, completely inconsistent with the expected 5--10 Myr.  The {\it green diamond} shows that forcing 2M1207B to lie on the 5--10 Myr isochrone, for its observed \lbol, requires \teff $\sim$ 1000K: 600K cooler than implied by its spectrum and colors.  Conversely, the {\it red diamond} shows that forcing 2M1207B to lie on the 5--10 Myr isochrone, at the derived \teff $\sim$ 1600K, requires an \lbol nearly an order of magnitude larger than observed.  See \S7.6. }

\plotone{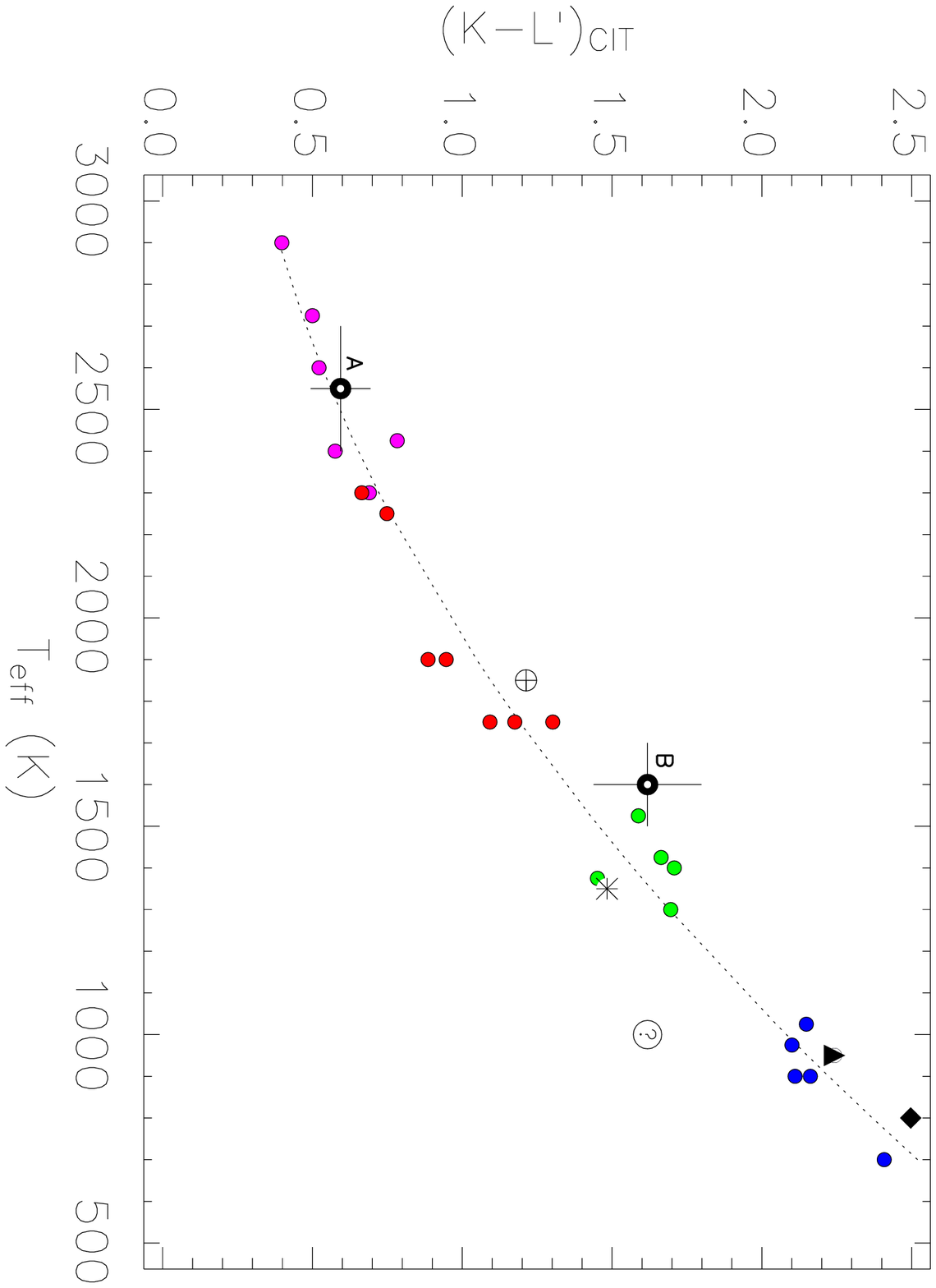}
\figcaption{\label{fig16} Empirical $K$-$L'$ versus \teff relationship in field M, L and T dwarfs.  The dwarfs are shown by {\it filled circles}: mid-to-late M (M6--M9.5) in {\it magenta}; early-to-mid L (L0--L5.5) in {\it red}; late L to early T (L6--T3) in {\it green}; and mid-to-late T (T4--T9) in {\it blue}.  The photometry errors for the field dwarfs are generally comparable to the symbol sizes.  The \teff shown assume a median age of 3 Gyr.  A few well-known dwarfs with independent age estimates are shown by special symbols: GD 165B (L3, {\it circled cross-hairs}), Gl 584C ({L8, \it asterisk}), Gl 229B (T6, {\it filled triangle}), and Gl 570D (T8, {\it filled diamond}).  The field dwarfs form a tight monotonically increasing sequence in $K$-$L'$ with decreasing \teff; the {\it dotted line} shows a second-order polynomial fit to the data to guide the eye.  Finally, 2M1207A and B are also plotted, as {\it bullseyes}, at their observed $K$-$L'$ and the \teff {\it derived} from our spectral and color analysis.  The position of both sources is fully consistent with the empirical field dwarf sequence.  The {\it circled question mark} denotes the position of 2M1207B if its \teff were actually 1000K, as required to match its \lbol but 600K cooler than we derive; this would clearly make it a significant outlier from the field dwarf color-\teff sequence shown.  See \S 8.3.    }

\clearpage

\begin{landscape}
\begin{deluxetable}{lccccccccccccc}
\tablecaption{\label{tab1}Observed and Derived Properties of 2MASS1207-3932 A \& B}
\tablewidth{0pt}
\tabletypesize{\scriptsize}
\tablehead{
\colhead{name} &
\colhead{J$_{2MASS}$\tablenotemark{a}} &
\colhead{H$_{2MASS}$\tablenotemark{b}} &
\colhead{K$_{s,2MASS}$\tablenotemark{b}} &
\colhead{L$'$\tablenotemark{b}} &
\colhead{sep\tablenotemark{a}} &
\colhead{PA\tablenotemark{a}} &
\colhead{SpT\tablenotemark{a}} &
\colhead{dist\tablenotemark{c}} &
\colhead{Age\tablenotemark{b}}  &
\colhead{log (L/\lsun)\tablenotemark{c}}  &
\colhead{\teff\tablenotemark{a}}  &
\colhead{Mass\tablenotemark{a}}  &\\
 &(mag) &(mag) &(mag) &(mag) & (mas) & ($^{\circ}$) & & (pc) & (Myr) &  & (K)& (M$_{Jup}$) \\}   
                            
\startdata

2M1207A & 13.00$\pm$0.03 & 12.39$\pm$0.03 & 11.95$\pm$0.03 & 11.38$\pm$0.10 & & & M8 & 53$\pm$6 & 5--10 & -2.68$\pm$0.12 & 2550$\pm$150 & 20--30\\
2M1207B & 20.0$\pm$0.2 & 18.09$\pm$0.21 & 16.93$\pm$0.11 & 15.28$\pm$0.14 & 769$\pm$10 & 125.6$\pm$0.7 & mid-to-late L & & & -4.72$\pm$0.14 & 1600$\pm$100 & 6--10\\

\enddata

\tablenotetext{a}{This work}
\tablenotetext{b}{Chauvin et al.(2004)}
\tablenotetext{c}{Mamajek (2005)}

\end{deluxetable}
\end{landscape}

\end{document}